\documentclass[journal]{IEEEtai}

\usepackage[colorlinks,urlcolor=blue,linkcolor=blue,citecolor=blue]{hyperref}

\usepackage{color,array}

\usepackage{graphicx}

\usepackage{booktabs}   
\usepackage{pifont}      
\usepackage{graphicx}
\usepackage[ruled,vlined,linesnumbered]{algorithm2e}
\usepackage{amsmath,amssymb,amsfonts}
\usepackage{todonotes}
\usepackage{soul}
\usepackage{balance}
\usepackage{hyperref}

\usepackage{tikz,pgfplots}
\pgfplotsset{compat=1.18}
\usetikzlibrary{arrows.meta,positioning,fit,shapes,calc}

\begin{document}





\title{AI-driven Predictive Shard Allocation for Scalable Next Generation Blockchains}

\author{
M. Zeeshan Haider, 
Tayyaba Noreen, 
M. D. Assunção, 
Kaiwen Zhang
\thanks{All authors are with the Department of Software Engineering, École de technologie supérieure (ÉTS), Montreal, Canada. 
Emails: \{mzeeshan.haider.1, tayyaba.noreen.1\}@ens.etsmtl.ca, Marcos.Dias-De-Assuncao@etsmtl.ca, kaiwen.zhang@etsmtl.ca.}
\thanks{Manuscript submitted on November 12, 2025, for review. 
This research was supported by the École de technologie supérieure (ÉTS) Research Fund and the institutional graduate research support program. 
The authors gratefully acknowledge the computational resources and infrastructure provided by the LASI lab at ÉTS.}

}
\maketitle

\begin{abstract}
Sharding has emerged as a key technique to address blockchain scalability by partitioning the ledger into multiple shards that process transactions in parallel. Although this approach improves throughput, static or heuristic shard allocation often leads to workload skew, congestion, and excessive cross-shard communication diminishing the scalability benefits of sharding. To overcome these challenges, we propose the Predictive Shard Allocation Protocol (PSAP), a dynamic and intelligent allocation framework that proactively assigns accounts and transactions to shards based on workload forecasts. PSAP integrates a Temporal Workload Forecasting (TWF) model with a safety-constrained reinforcement learning (Safe-PPO) controller, jointly enabling multi-block-ahead prediction and adaptive shard reconfiguration. The protocol enforces deterministic inference across validators through a synchronized quantized runtime and a safety gate that limits stake concentration, migration gas, and utilization thresholds. By anticipating hotspot formation and executing bounded, atomic migrations, PSAP achieves stable load balance while preserving Byzantine safety. Experimental evaluation on heterogeneous datasets, including Ethereum, NEAR, and Hyperledger Fabric mapped via address-clustering heuristics, demonstrates up to 2× throughput improvement, 35\% lower latency, and 20\% reduced cross-shard overhead compared to existing dynamic sharding baselines. These results confirm that predictive, deterministic, and security-aware shard allocation is a promising direction for next-generation scalable blockchain systems.

\end{abstract}

\begin{IEEEImpStatement}
Our paper introduces the \textbf{Predictive Shard Allocation Protocol (PSAP)}, an AI-driven framework that transforms blockchain scalability into a continuous, predictive decision process. Unlike heuristic or static methods, PSAP integrates Temporal Workload Forecasting (TWF) with a Safe-PPO controller to proactively and safely reallocate resources in real time. A key contribution is the \textbf{Deterministic Machine Learning Execution Layer (DMEL)}, which ensures reproducible and verifiable inference across heterogeneous validators while enforcing constraints on stake, gas, and latency. This closes a major gap between learning-based intelligent control and deterministic execution in decentralized environments. Experiments show that PSAP improves throughput by up to 2$\times$ and reduces latency by 35\% while preserving full Byzantine-fault tolerance, demonstrating its broader potential for secure, autonomous digital infrastructure.

\end{IEEEImpStatement}


\section{Introduction}

\IEEEPARstart{B}{lockchain} networks such as Bitcoin and Ethereum face inherent scalability limitations. Bitcoin is limited to processing roughly seven Transactions Per Second (TPS), while Ethereum averages about fifteen TPS, both are far below centralized payment systems such as Visa, which regularly handle over 1,700 TPS \cite{yi2024data}. This gap arises from the design principle that every node in a blockchain must process and validate every transaction, ensuring security and decentralization at the expense of throughput~\cite{yuan2024secure}. As adoption grows, this bottleneck threatens the ability of blockchains to support high-volume applications~\cite{huang2025blockemulator}. Sharding has been proposed as a promising first-layer scalability solution that divides the blockchain into multiple smaller partitions, or shards, each responsible for processing a fraction of the total transactions \cite{yi2024data}. By distributing transactions across multiple shards for parallel execution, sharding provides the potential to achieve near-linear scalability. Ideally, increasing the number of shards by a factor of two would yield a proportional increase in throughput, approaching a twofold improvement~\cite{zhen2024dynamic}. Ethereum 2.0 incorporates this idea with plans to introduce 64 shards, which could increase capacity to tens of thousands of TPS. However, practical deployment of sharded blockchains has shown that ideal scaling is difficult to achieve in practice. Performance degradation arises from challenges such as workload imbalance across shards and the high cost of coordinating cross-shard transactions~\cite{huang2022brokerchain}.

One major obstacle is that naive shard assignment strategies often rely on random distribution or hashing, which do not consider workload heterogeneity~\cite{li2024reputation}, resulting in uneven load distribution, where some shards become heavily congested with transaction requests while others remain underutilized. Overloaded shards accumulate long queues and slow down overall system throughput, while idle shards waste resources \cite{akshatha2023mqtt}. The problem is compounded by cross-shard transactions, which require communication between shards when a transaction involves accounts or data split across multiple partitions~\cite{li2024spring}. Coordination mechanisms, such as atomic commit protocols, introduce additional latency and network overhead. Prior studies indicate that such inefficiencies can reduce throughput by more than 30\% compared to ideal scaling~\cite{nguyen2025blockchain}. 

Adaptive shard allocation is required to manage workload fluctuations and network conditions in real time \cite{jiang2024resource}. By leveraging blockchain data such as transaction rates and account activity, machine learning can forecast demand, predict hot-spots, and proactively rebalance load. Predictive shard allocation can reduce cross-shard communication by clustering frequently interacting accounts, lowering overhead and latency \cite{henry2018blockchain}. PSAP implements intelligent algorithms for dynamic shard assignment to enhance performance and security, which integrates timeseries prediction and reinforcement learning to optimize shard configuration continuously. Our key contributions include:

\begin{enumerate}
    \item \textbf{Temporal workload forecasting:} A hybrid predictor combining deep recurrence and statistical signals to forecast per-shard load several blocks ahead for proactive congestion control.
    
    \item \textbf{Intelligent shard reallocation:} A two-layer mechanism that leverages predicted loads and reinforcement learning feedback to dynamically migrate accounts and reschedule transactions, optimizing throughput and latency in real time.
    
    \item \textbf{Integrated shard allocation:} A unified framework that fuses workload forecasting with adaptive reallocation, providing a closed-loop control system for continuously optimizing shard configuration under dynamic conditions.
\end{enumerate}

The remaining sections are organized as follows. Section~\ref{sec:background} presents the background and related work. Section~\ref{sec:protocol} introduces our proposed protocol, followed by Section~\ref{sec:implementation}, which describes the implementation. Section~\ref{sec:evaluation} reports the evaluation results, and Section~\ref{sec:conclusion} concludes the paper.

\section{Background and Related Work}
\label{sec:background}
Blockchain sharding divides the blockchain into smaller subsets of transactions, or shards, that operate in parallel. This section provides a structured overview of prior research, tracing the evolution from early throughput-oriented designs to predictive and reinforcement learning–based allocation models. We categorize the literature into five key themes: (1) foundational sharding systems, (2) range-based and deterministic placement, (3) heuristic and workload-driven allocation, (4) cross-shard communication and security, and (5) predictive and reinforcement learning–based sharding.

\subsection{Foundational Sharding Systems}

Pioneering sharding systems, such as Elastico~\cite{luu2016elastico} and OmniLedger~\cite{kokoris2018omniledger}, have demonstrated that throughput-proportional secure sharded ledgers are feasible. Research in this field has evolved in two main directions. The first investigates how to assign states and validators to shards, while the second focuses on efficiently routing cross-shard transactions without compromising security. Approaches such as RapidChain, QuorumChain, Ethereum~2.0 (phase~0), and NEAR's Nightshade~2.0 map accounts or validators to shards using hashing or static ranges, periodically rotating validator committees once per epoch to limit adversarial influence \cite{zamani2018rapidchain,buterin2020eth2,skidanov2024nightshade}.

\subsection{Range-Based and Deterministic Shard Placement}
While these epochic rotations maintain classical Byzantine-fault thresholds, the underlying state layout remains largely fixed. As a result, persistent workload skew carries over from one epoch to the next, leading to a throughput collapse when activity concentrates on a few contracts or accounts. Empirical replays of Ethereum and Zilliqa workloads confirm that a single busy DeFi pool can reduce throughput by over 30\% and double or triple transaction latency \cite{li2023lbchain,xiao2024spring}. Each rotation adds protocol overhead, consuming 3–5\% bandwidth in NEAR mainnet \cite{skidanov2024nightshade}. The deterministic range and partitioning of UTXO reduce the complexity of the proof but are brittle: nearly half of daily transfers occur in just one-eighth of its address space, resulting in 27\% throughput reduction~\cite{zilliqaMetrics2024}. Range-based schemes also lack flexibility, requiring costly re-sharding and checkpointing in systems such as Fabric 2.x~\cite{fabricShards2023}.

\subsection{Heuristic and Workload-Driven Allocation}

To overcome earlier limitations, subsequent work adopted reactive, heuristic-based reallocation approaches. Graph clustering schemes, such as BrokerChain, build transaction graphs every few blocks and apply community detection to co-locate frequently interacting accounts \cite{huang2024brokerchain}, thereby reducing cross-shard edges but incurring heavy rebuild and proof costs. Deterministic placement with zero-knowledge receipts, as in Shard Scheduler, still requires substantial state movement \cite{krol2021shardscheduler}. Threshold-triggered methods migrate once gas use exceeds a limit (Garet \cite{woo2020garet}) or split/merge shards dynamically (DynaShard \cite{liu2025dynashard}). Still, all remain reactive, intervening only after congestion appears and often inducing transient instability. 

\subsection{Predictive and Statistical Forecasting Approaches}

Recent studies have explored prediction-driven shard allocation to improve scalability. LB-Chain employs LSTMs to predict account activity and migrate top senders, reducing confirmation delays by up to 90\%~\cite{li2023lbchain}. HMMDShard and ShardBalance use statistical models like HMM and ARIMA for workload forecasting~\cite{huang2024hmmdshard,kim2023shardbalance}, but rely on fixed parameters and lack adaptability under dynamic or adversarial conditions. RL-based systems such as SkyChain, Spring, and AERO~\cite{zhang2020skychain,xiao2024spring,song2025aero} enhance throughput using deep Q-learning or DRL clustering, yet most operate offline with periodic retraining, lack deterministic inference, and provide no formal safety or adversarial guarantees. These gaps underscore the need for an online, secure, and self-adaptive predictive framework addressed by PSAP in this work.

\begin{table*}[t]
\centering
\caption{Comparative analysis of recent sharding frameworks (forecasting, RL, safety, adaptability).}
\label{tab:comparison}
\renewcommand{\arraystretch}{1.05}
\scriptsize
\setlength{\tabcolsep}{3pt}
\resizebox{\textwidth}{!}{%
\begin{tabular}{lcccccccc}
\toprule
\textbf{Framework} &
\textbf{Forecast} &
\textbf{Shard Allocation} &
\textbf{Joint mig.} &
\textbf{Safety} &
\textbf{Updates} &
\textbf{Heterog.} &
\textbf{Adv. res.} \\
\midrule
Nightshade 2.0~\cite{skidanov2024nightshade} &
\ding{55} & \ding{55}  & \ding{55}  & \ding{55}  & Epoch & \ding{55}  & \ding{55}  \\

LB-Chain~\cite{li2023lbchain} &
\checkmark\ (epochic $\leq$1) & None & \ding{55}  & \ding{55}  & Epoch & Eth & \ding{55}  \\

DynaShard~\cite{liu2025dynashard} &
\ding{55}  & \ding{55}  & Split/Merge & \ding{55}  & Periodic & Priv & \ding{55}  \\

AI-Shard~\cite{zhao2024aishard} &
\checkmark\ ($\leq$4) & \ding{55}  & Partial & -- & Epochic & Synth & \ding{55}  \\

AERO~\cite{song2025aero} &
\ding{55} &  static & Partial & \ding{55}  & Offline & Eth & \ding{55}  \\

HydraChain~\cite{ma2025hydrachain} &
Adaptive count & \ding{55}  & \ding{55} & \ding{55}  & Periodic & Pub & \ding{55}  \\

\midrule
\textbf{PSAP (ours)} &
\textbf{\checkmark\ (multi $H{\leq}32$)} &
\textbf{Safe-PPO (fine)} &
\textbf{\checkmark\ (timing\&scope)} &
\textbf{\checkmark\ (gas/stake/util.)} &
\textbf{Per-blk} &
\textbf{Pub+Priv} &
\textbf{\checkmark} \\
\bottomrule
\end{tabular}}
\end{table*}

\subsection{Cross-Shard Communication and Security}

Cross-shard communication and security remain critical challenges across all allocation strategies. Protocols such as Atomix, two-phase commit variants, and Rooted-Graph Placement aim to reduce inter-shard message complexity~\cite{kokoris2018omniledger,ren2021rooted}, while Wormhole demonstrates that no protocol simultaneously achieves both efficient self-balancing and full interoperability~\cite{ghorban2022wormhole}. From a security standpoint, random validator selection and bounded migration are essential to prevent shard-takeover attacks~\cite{wang2023sok}. However, most dynamic schemes still trade off between scalability, coordination overhead, and adversarial resilience. Existing approaches rarely integrate horizon-based forecasting, continuous online refinement, and explicit security guarantees within one unified framework leaving a clear gap that PSAP addresses through predictive, safety-constrained cross-shard coordination.

\subsection{Reinforcement Learning–Based Shard Adaptation}

Existing research on adaptive sharding has largely focused on forecasting rather than closed-loop control. Most frameworks, such as AI-Shard~\cite{zhao2024aishard}, OBCS~\cite{zhuang2023obcs}, and HydraChain~\cite{ma2025hydrachain}, employ workload prediction but lack reinforcement learning–driven decision mechanisms. Their predictive horizons are short (typically $\leq4$ blocks) or limited to coarse shard-count adjustments, offering no fine-grained or continuous adaptation. LB-Chain~\cite{li2023lbchain} performs epoch-level LSTM forecasting without online feedback, while AERO~\cite{song2025aero} retrains DRL agents offline, applying periodic policy updates that respond slowly to dynamic workloads. Thus, no prior work has demonstrated a unified framework that jointly integrates real-time forecasting with an online reinforcement learning allocator.

In contrast, PSAP introduces a \emph{multi-block, per-block} control loop that fuses Temporal Workload Forecasting (TWF) with a Safe-PPO reinforcement controller, enabling proactive and fine-grained account-level reallocation. Unlike existing methods that overlook determinism and safety, PSAP enforces a \emph{safety gate} that bounds Byzantine stake, migration gas, and utilization before on-chain execution, ensuring secure and consistent decisions across validators. Table~\ref{tab:comparison} highlights these algorithmic differences. Hence, PSAP fills a critical research gap by providing the first continuously adaptive, safety-constrained, and deterministic reinforcement learning–based framework for predictive shard allocation.

\section{Predictive Shard Allocation Protocol (PSAP)}
\label{sec:protocol}

In this section, we provide a detailed description of our proposed protocol. We first outline the system model and assumptions, then present the predictive algorithms: a time series forecasting module using TWF and a RL based shard adjustment. We provide full equations for the predictive models and the RL formulation, followed by the shard reallocation algorithm that uses these predictions.

\subsection{Design Goals and Assumptions}
\label{subsec:design}

PSAP operates in a sharded blockchain, where each shard maintains Byzantine Fault Tolerance (BFT) consensus and a global commit layer ensures cross-shard finality. The protocol predicts upcoming load hot-spots and rebalances state proactively while preserving security and minimizing communication overhead. We model a permissionless system with shards $\mathcal{S}=\{S_1,\ldots,S_K\}$, each tolerating $f<\tfrac{1}{3}$ Byzantine validators. Time advances in slots of $\Delta_{\mathrm{blk}}$ seconds, and PSAP forecasts up to $H\!\le\!32$ future slots corresponding to 30–60 seconds for typical block intervals to balance forecasting accuracy with computational and synchronization limits. Validators in each shard $s$ at height $t$ export $(x_s(t), g_s(t), \sigma_s(t))$, representing transactions, gas usage, and stake share for the LSTM forecaster and Safe Proximal Policy Optimization (Safe-PPO) controller.

We assume a polynomial-time adversary who can corrupt at most
$\beta N$ validators across the whole network, with
$\beta < \tfrac13$ so that no single shard ever exceeds the
Byzantine-stake threshold. In addition to compromising validators, the adversary can \emph{steer} traffic by creating Sybil clients and manipulating fee queues to create artificial hot-sp1ots, and can try to manipulate honest validators to concentrate stake in a target shard. We consider that standard cryptographic primitives such as hash, BLS signatures and VRF randomness are secure \cite{ifrim2024systematic}. The network is partially synchronous with bounded delay~$\Delta$, where the adversary can schedule but cannot drop or tamper with messages. Under these assumptions, the BFT protocol ensures safety/liveness, and PSAP's safety gate blocks migrations that would raise a shard's Byzantine stake above~$\tfrac13$.

\begin{figure*}[t]
  \centering
  \includegraphics[width=\textwidth]{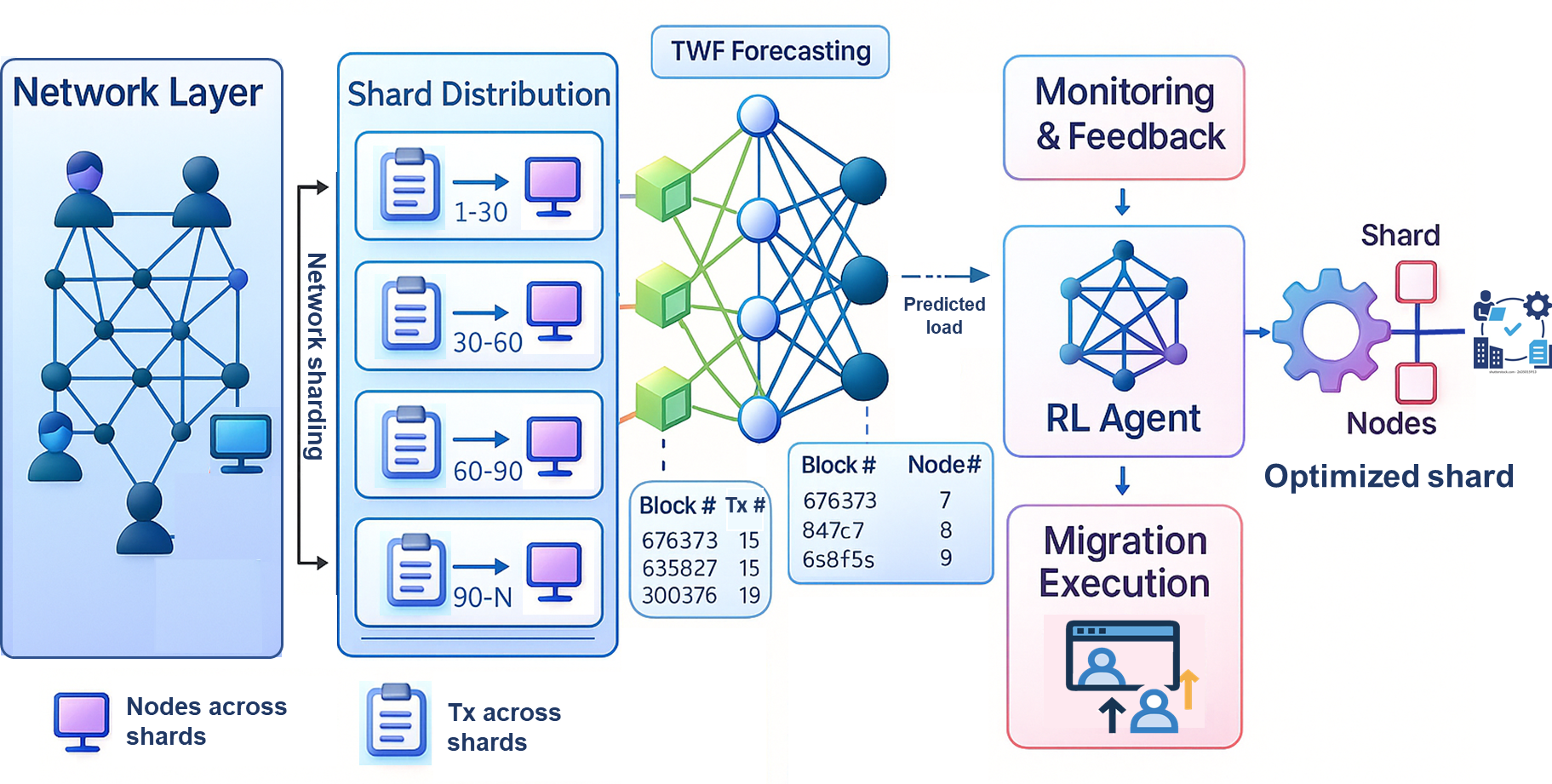}
  \caption{Overall architecture of the PSAP protocol.}
  \label{fig:threat-workflow}
\end{figure*}


Our design adheres to three key principles: predictive load balancing for scalable throughput, security against Byzantine takeover, and bounded migration overhead. It also enforces fairness by limiting fee dispersion and maintaining stable performance under dynamic workloads. The threat model assumes up to one-third of colluding validators and strategic attacks, such as hotspot creation or stake bribery.
Central to defending against both is the \emph{safety gate} placed between the RL agent's output and the execution layer, which checks every candidate migration plan against strict constraints: it must respect the $\tfrac{1}{3}$ stake bound, the 2\% gas ceiling, the 0.5\% account-move limit, and may not push any shard above 0.75 utilisation in expectation. Plans violating constraints are clipped or rejected, restricting the allocator to valid actions. A TWF forecaster provides an $H$-block look-ahead. Safe-PPO maps predictions to migration batches that minimize imbalance and gas cost. The on-chain contract then applies them atomically with dual inclusion to ensure liveness and rollback safety.

\subsection{System Model \& Notation}

We model the sharded blockchain as a discrete-time system with $K$ shards 
$\mathcal{S}=\{S_{1},\dots,S_{K}\}$, each running BFT consensus with at most 
$f_{s}<\tfrac{1}{3}n_{s}$ Byzantine validators. Shard capacity $C_{s}$ is 
measured in gas or transactions per second. Accounts 
$\mathcal{A}=\{a_{1},\dots,a_{N}\}$ are mapped to shards by an allocation 
function $\pi(t):\mathcal{A}\to\mathcal{S}$. A transaction 
$\tau=\langle \text{src},\text{dst},\text{val},\text{gas}\rangle$ is 
\emph{intra-shard} if $\pi(\text{src})=\pi(\text{dst})$, else 
\emph{cross-shard}, with ratio
\[
  \rho(t)=\frac{\#\{\tau\mid \pi(\text{src})\neq \pi(\text{dst})\}}{\#\{\tau\}}.
\]

Shard load is captured by transaction count $x_{s}(t)$, gas $g_{s}(t)$, and 
utilization $\mathrm{util}_{s}(t)=x_{s}(t)/C_{s}$, aggregated in 
$\mathbf{x}(t)$. The forecaster predicts 
$\hat{\mathbf{x}}(t+1{:}t+H)$ to guide migration up to $H$ blocks ahead. 
At each $t$, the RL allocator outputs a batch
\[
  a(t)=\{\langle a_{j},S_{p(j)},S_{q(j)}\rangle\}_{j=1}^{m(t)}, \quad m(t)\le m_{\max}.
\]
Cumulative migrations are 
$\mathcal{M}(T)=\sum_{t=1}^{T}m(t)$. Safety requires (i) Byzantine stake 
$<\tfrac{1}{3}$ per shard and (ii) migration gas $M(t)\le 0.02$ of block gas. 
Violations trigger a \emph{safety gate} that prunes entries. The system dynamics can thus be abstracted as
$\mathbf{x}(t+1)=F(\mathbf{x}(t),a(t),\zeta(t))$, where $\zeta(t)$
represents exogenous arrivals. The forecasting module minimizes expected
prediction error
$\min_{\theta}\mathbb{E}[\|\mathbf{x}(t+1)-\hat{\mathbf{x}}_{\theta}(t+1)\|^{2}]$,
while the allocator optimizes a constrained long-term reward
\[
  \max_{\pi}\;
  \mathbb{E}_{\pi}\!\Big[\sum_{t=0}^{\infty}\gamma^{t}R(\mathbf{x}(t),a(t))\Big],
  \quad \text{s.t. safety constraints.}
\]
This compact formalization provides the foundation for the TWF forecaster in Section~\ref{subsec:twf} and the RL allocator in
Section~\ref{subsec:rl}.

\begin{table}[!t]
  \caption{Key symbols and notation.}
  \label{tab:symbols}
  \centering
  \renewcommand{\arraystretch}{1.1}
  \begin{tabular}{p{1.1cm}p{6.3cm}}
    \toprule
    \textbf{Symbol} & \textbf{Meaning} \\ \midrule
    $K$ & Number of shards \\ 
    $S_{s}$ & Shard index $s\in\{1..K\}$ \\
    $V_{s}(t)$ & Validator set of shard $S_{s}$ at block $t$ \\
    $C_{s}$ & Capacity (gas or tx/s) of shard $S_{s}$ \\
    $\pi(t)$ & Account–to–shard mapping at block $t$ \\
    $x_{s}(t)$ & \# transactions executed by $S_{s}$ at block~$t$ \\
    $g_{s}(t)$ & Gas consumed by $S_{s}$ at block~$t$ \\
    $\mathrm{util}_{s}(t)$ & Utilisation of shard $S_{s}$ \\
    $\mathbf x(t)$ & Load vector $[x_{1},\dots,x_{K}]^{\!\top}$ \\
    $H$ & Forecast horizon (blocks) \\
    $a(t)$ & Migration action proposed at block~$t$ \\
    $I(t)$ & Imbalance metric (Eq.\,2) \\
    $M(t)$ & Migration-cost metric (Eq.\,3) \\
    $R(t)$ & RL reward (Eq.\,4) \\ \bottomrule
  \end{tabular}
\end{table}

\subsection{End-to-End Workflow}
\label{subsec:workflow}

The PSAP workflow (Fig.~\ref{fig:threat-workflow}, Alg.~\ref{alg:psap-init}) executes at each block cycle, where every validator runs the same pipeline and derives consistent actions from the shared system state without requiring additional consensus steps. The process starts with \textbf{metrics logging}. When a block $t$ is
finalised, every validator records the tuple
\begin{equation}
\langle x_{s}(t), g_{s}(t), \sigma_{s}(t) \rangle,
\end{equation}
for its shard $S_{s}$. A compact summary of these values, amounting to
only two bytes per metric per shard, is then gossiped across the
network. Next, the system performs \textbf{forecasting module}. Each validator uses
its local forecasting module to process a sliding window of shard metrics of length $W$,
\begin{equation}
\{\mathbf{x}(t-W+1), \dots, \mathbf{x}(t)\},
\end{equation}
and outputs an $H$-step prediction of the workload,
\begin{equation}
\hat{\mathbf{x}}(t+1 : t+H),
\end{equation}
which adds approximately $28$ ms of inference overhead on a 16-vCPU
node. Following this, the reinforcement learning decision stage is then executed locally by each validator, ensuring consistent shard adjustment without requiring a centralized controller. When validators migrate across shards, their RL models continue from synchronized policy parameters shared through periodic state updates, avoiding retraining or trusted third-party dependence. The PPO policy network takes as input the concatenated state vector and outputs a migration proposal, where each element specifies the relocation of an account $a_{j}$  from source shard $S_{p}$ to destination shard $S_{q}$.  Before these moves are committed, the proposals are passed through the \textbf{safety gate}, which enforces two critical constraints: 
(i) the post-migration Byzantine stake fraction in each shard must remain below one-third, and (ii) the total migration cost must not exceed 2\% of the block gas limit. 
If either condition is violated, the proposal $a(t)$ is pruned until it satisfies all safety constraints.

\begin{equation}
s(t) = [\mathbf{x}(t), \hat{\mathbf{x}}(t), \boldsymbol{\sigma}(t)],
\end{equation}

\begin{equation}
a(t) = \{\langle a_{j}, S_{p}, S_{q}\rangle\}_{j=1}^{m},
\end{equation}

\begin{equation}
M(t) \leq 0.02 \cdot G_{\text{block}},
\end{equation}
In the \textbf{migration commit} phase, all safe moves are aggregated into a 
\textsf{Mig\_Batch} that is ordered through the shard's BFT consensus at block $t+1$. Each batch carries a compact cross-shard proof, represented by one Merkle root per destination shard, which finalizes the updated mapping $\pi(t+1)$. Every migration is anchored to the source shard height $t$, and tags are attached to ensure idempotent replay across forks or retries. The protocol is designed so that the commit phase executes entirely within the block interval, without requiring any additional consensus rounds. This guarantees that reallocation remains atomic, consistent, and rollback-safe, while preserving liveness under partially synchronous 
network conditions.

\subsection{Temporal Workload Forecasting (TWF)}
\label{subsec:twf}

PSAP anticipates congestion through proactive account reallocation, forecasting future load hot-spots in advance, rather than relying on reactive balancing schemes that trigger migrations only after shards become congested. It employs a TWF engine based on a lightweight single-layer LSTM,  cross-shard correlated transaction streams. The forecaster outputs multi-step load predictions that guide the RL allocator. At every block height $t$, validators assemble an input sequence of
length $W$, denoted
\[
\mathcal{W}_{t}=\{\mathbf{f}(\tau)\}_{\tau=t-W+1}^{t},
\]
where each feature vector $\mathbf f(\tau)$ includes the per-shard
transaction counts $\mathbf{x}(\tau)=[x_{1}(\tau),\ldots,x_{K}(\tau)]^{\top}$,
the total gas expenditure $g(\tau)$, and a periodic encoding of the hour
of the day. Specifically, the time-of-day features are constructed as
\[
\mathrm{hour\_sin}(\tau)=\sin\!\left(\tfrac{2\pi}{24}\cdot \mathrm{hour}(\tau)\right), 
\quad
\mathrm{hour\_cos}(\tau)=\cos\!\left(\tfrac{2\pi}{24}\cdot \mathrm{hour}(\tau)\right),
\]
which encode diurnal transaction cycles observed in both public and
enterprise deployments. The complete feature vector therefore takes the
form
\[
\mathbf f(\tau)=
\bigl[\,\mathbf x(\tau),\,g(\tau),\,\mathrm{hour\_sin}(\tau),\,\mathrm{hour\_cos}(\tau)\,\bigr]^{\top}.
\]
To stabilize training, features use an exponential moving average  for adaptive streaming normalization. The forecaster processes the input sequence sequentially according to
\[
(h_{\tau},c_{\tau})=\mathrm{LSTM}\!\left(\mathbf f(\tau),\,h_{\tau-1},\,c_{\tau-1}\right),
\]
where $h_{\tau},c_{\tau}\in\mathbb{R}^{d_{h}}$ denote the hidden and
cell states with dimension $d_{h}=64$. After unrolling over the window
$\mathcal{W}_{t}$, the terminal hidden state $h_{t}$ is projected
linearly into the forecasted load vector
\[
\hat{\mathbf x}(t+1)=W_{\mathrm{out}}h_{t}+b_{\mathrm{out}}.
\]
For horizons $H>1$, training employs teacher forcing, while inference
adopts autoregressive rollout:
\[
\hat{\mathbf x}(t+j)=F_{\theta}\!\left(\hat{\mathbf x}(t+j-1),h_{t+j-1}\right), 
\quad j=2,\ldots,H,
\]
where $F_{\theta}$ denotes the parameterized recurrent mapping. The
learning objective is the minimization of normalized mean-squared error,
\[
\mathcal{L}(\theta)=\frac{1}{KH}\sum_{s=1}^{K}\sum_{j=1}^{H}
\frac{\bigl(x_{s}(t+j)-\hat{x}_{s}(t+j)\bigr)^{2}}{\sigma_{s}^{2}},
\]
where $\sigma_{s}^{2}$ is the variance of shard $s$, ensuring balanced
accuracy across heterogeneous load distributions. The training corpus combined traces from Ethereum Sepolia, NEAR, Zilliqa, and Hyperledger Fabric, spanning both permissionless and permissioned settings. 
Data were partitioned into training, validation, and test splits. 
Models were optimized with Adam using a learning rate $\eta=10^{-3}$ and cosine decay, together with regularization techniques including dropout ($p=0.2$), weight decay ($\lambda=10^{-5}$), and early stopping. 
Inference remains lightweight, executing well within the $100$\,ms per-block budget, ensuring compatibility with high-throughput chains. 
To quantify uncertainty, Monte Carlo dropout with $N=10$ samples is applied, producing predictive variances $\sigma^{2}_{s}(t+j)$. 
The allocation agent then hedges using the upper confidence bound
\[
\tilde{x}_{s}(t+j) = \hat{x}_{s}(t+j) + 1.96\sigma_{s}(t+j),
\]
which ensures with high probability that shard load will not exceed the forecast safe thresholds. 
This conservative adjustment prevents over-commitment and strengthens RL-driven allocation. 
By encoding diurnal cycles and incorporating uncertainty estimates, the TWF engine enables PSAP to forecast multiple blocks ahead, supporting proactive account reallocation. 
This preemptive approach contrasts with reactive balancing by addressing congestion before it materializes, thereby reducing coordination overhead.

\begin{algorithm}[!t]
\caption{PSAP per-block control loop at validator $v$}
\label{alg:psap-init}
\small
\SetKwInput{KwIn}{Input}  
\SetKwInput{KwOut}{Output}
\SetKwInput{KwPar}{Params}

\KwIn{$\mathcal{W}=\{\mathbf{x}(t\!-\!W{:}t)\}$ , stake vector $\boldsymbol{\sigma}(t)$}
\KwOut{Authenticated \textsf{MIG\_BATCH} and applied migrations}
\KwPar{budget $B$, cooldown $C$, quorum $q$, timeout $\tau$}

\ForEach{shard $S_s$}{
  pull latest stats $\langle x_s(t), g_s(t)\rangle$; append to $\mathcal{W}$ and prune to $W$\;
}

$\hat{\mathbf{x}},\,\mathbf{u}\gets\textsc{LSTM\_InferWithUnc}(\mathcal{W})$\;  
$a\gets\textsc{RL\_Policy}\!\bigl[\mathbf{x}(t),\hat{\mathbf{x}},\mathbf{u},\boldsymbol{\sigma}(t)\bigr]$\;

$a\gets\textsc{SafetyGate}\!\bigl(a,\boldsymbol{\sigma}(t)\bigr)$ 

$a\gets\textsc{ApplyBudget}(a,B)$; $a\gets\textsc{EnforceCooldown}(a,C)$\;

\If{$a=\varnothing$}{\Return} 

$(\mathsf{hdr},\mathsf{mov})\gets\textsc{BuildBatch}(a)$; $\mathsf{hdr.parent}\gets\textsc{Hash}(\text{last\_finalized})$\;
$\mathsf{sig}\gets\textsc{BLS\_AggSign}(\mathsf{hdr})$; 
\textsc{Broadcast}(\textsf{MIG\_BATCH}$\langle\mathsf{hdr},\mathsf{mov},\mathsf{sig}\rangle$)\;

$\mathcal{R}\gets\textsc{CollectVotes}(\mathsf{hdr},\tau)$\;
\If{$\textsc{Weight}(\mathcal{R})<q$}{\Return} 

$\mathsf{cert}\gets\textsc{CommitAtFinality}(\mathsf{hdr},\mathcal{R})$\;
\If{$\mathsf{cert}=\bot$}{\Return}  

\If{\textsc{ExecuteMoves}($\mathsf{mov}$) fails}{
  \textsc{RollbackMoves}($\mathsf{mov}$); \Return
}

\textsc{UpdateState}($\boldsymbol{\sigma}(t{+}1)$); \textsc{AuditLog}$(t,|\mathsf{mov}|,\mathbf{u},\mathsf{cert})$\;
\Return
\end{algorithm}

\subsection{RL Allocation Agent}
\label{subsec:rl}
Proximal Policy Optimization (PPO) is a policy-gradient method that stabilizes training by clipping updates, making it effective for real-time decision making. However, PPO optimizes 
only for reward and ignores safety constraints. Applied to shard reallocation, this may exceed migration budgets, gas ceilings, or Byzantine stake thresholds. PSAP therefore adopts Safe-PPO, which augments PPO with constraint-aware optimization 
via cost critics and adaptive penalties. It enforces the safety gate: Byzantine stake $<1/3$ per 
shard, migration gas $<2\%$ of block capacity, account moves $<0.5\%$, and utilization $\leq0.75$. 
Unsafe actions are pruned or penalized, ensuring throughput and security. The allocation problem 
is modeled as a Markov Decision Process (MDP).
 At block $t$,
the agent observes the state:
\[
  s_{t} = \bigl[\,
    \mathbf x(t),\;
    \hat{\mathbf x}(t{:}t+H),\;
    \sigma_{1{:}K}(t),\;
    \mathrm{var}\bigl(\hat{\mathbf x}\bigr)
  \bigr],
\]
The state has dimension $2K(H{+}1)+K$, capturing current load, $H$-step forecasts, and predictive uncertainty. Explicitly encoding variance enables conservative actions under high uncertainty, crucial for volatile workloads like flash-loan arbitrage. The action $a_{t}$ consists of up to $m_{\max}=100$ account migrations,
\[
  a_{t} = \bigl\{\langle a_{j},\,S_{p(j)},\,S_{q(j)}\rangle\bigr\}_{j=1}^{m_{t}}, 
  \quad m_{t}\le m_{\max}.
\]
Given the combinatorial action space, we adopt a Pointer-Net head \cite{yang2025generative} to rank accounts by load impact and migration cost, scaling linearly with accounts for validator deployment. The reward balances throughput, fairness, and efficiency:
\[
  r_{t} = \mathrm{TPS}(t) - \alpha I(t) - \beta M(t), \quad \alpha=\beta=0.5,
\]
where $I(t)$ is shard imbalance and $M(t)$ migration cost. Coefficients tune trade-offs between aggressive and conservative allocation. A hard cost signal
\[
  c_{t} = \max(M(t)-0.02,\,0),
\]
captures violations of the 2\% migration gas budget. This dual reward-cost design makes the policy both incentive-compatible and constraint-aware. The optimization follows Lagrangian-constrained PPO, with cost critic $V^{c}_{\psi}$ and adaptive multiplier $\lambda$:
\begin{align}
  \max_{\theta}\;&
    \mathbb{E}_{\pi_\theta}\!\Bigl[\tfrac{1}{1-\gamma}\sum_{t}\gamma^{t}(r_{t}-\lambda c_{t})\Bigr], \\
  \text{s.t.}\;& \mathbb{E}[c_{t}] \le 0.
\end{align}
Here $\lambda$ penalizes unsafe actions via gradient feedback, ensuring feasible convergence unlike vanilla PPO. Policy and value share a 128-unit MLP with SELU activations, and the pointer head uses two attention layers for account selection. After policy inference, validators apply a deterministic safety gate:
\[
  \forall S_{q}:\;\sigma_{q}^{\mathrm{Byz}}(t+1) < \tfrac{1}{3}, \quad M(t)\le0.02.
\]
If violated, worst-impact moves are pruned until both conditions hold. This guarantees verifiable safety without extra consensus rounds. The agent is lightweight: a forward pass takes $\sim$15\,ms, and one PPO update every $N=20$ blocks costs 42\,ms on CPU, well within a 2\,s slot. It is pretrained on 5M blocks with TWF features for initialization, then fine-tuned online to handle \emph{concept drift}. Empirically, it re-stabilizes within $\sim$300 blocks ($<$10 min at 2\,s/block).



\subsection{Migration Execution Logic}
\label{subsec:migration}

After passing the safety gate, reallocations are materialized on-chain via a \textsf{Mig\_Batch} transaction. Generated by the source shard $S_p$ and signed by $2f_p{+}1$ validators, it is inserted as the first user transaction of block $t{+}1$. Destination shards verify the BLS aggregate, Merkle root, and local height $\ge t{+}1$, then pause incoming accounts, validate, and append an \textsf{Apply\_Mig} record. From block $t{+}2$, accounts are treated as native, and cross-shard transfers use updated tags. Dual inclusion across source and destination ensures atomic commit without two-phase protocols. To ensure liveness under partial failures, PSAP employs rollback. If a destination shard fails to confirm \textsf{Apply\_Mig} within $\tau_{\text{timeout}}=3$ blocks, the source shard issues \textsf{Cancel\_Mig}, reverting moves, restoring balances, and compensating receipts. The safety gate budgets gas so either \textsf{Mig\_Batch} or \textsf{Cancel\_Mig} fits within the 2\% block-gas cap. This guarantees atomic cross-shard migrations without extra consensus rounds.

\begin{algorithm}[t]
\caption{Safe-PPO update executed every $N$ blocks}
\label{alg:safe-ppo}
\scriptsize
\DontPrintSemicolon
\setlength{\algomargin}{0.5em}
\SetKwInput{KwIn}{Input}\SetKwInput{KwOut}{Output}\SetKwInput{KwPar}{Params}

\KwIn{policy $\pi_{\theta}$, reward value $V_{\phi}$, cost value $V^{c}_{\psi}$, multiplier $\lambda$, buffer $\mathcal{D}$}
\KwOut{Updated $\theta,\phi,\psi,\lambda$}
\KwPar{epochs $K_{\text{ep}}$, minibatch $M$, clip $\epsilon$, entropy $\beta$, lrs $\eta_{\theta},\eta_{\phi},\eta_{\psi}$, KL cap $\mathrm{KL}_{\max}$}

\textbf{Rollout:} run $\pi_{\theta}$ for $N$ blocks; collect $B$ tuples $(s_t,a_t,r_t,c_t,s_{t+1})\!\in\!\mathcal{D}$.\;

\textbf{GAE:} compute $\hat{A}^{r},\hat{A}^{c}$ with $\lambda_{\text{GAE}}{=}0.95$ and targets $R_t^{r},R_t^{c}$ for $V_{\phi},V^{c}_{\psi}$.\;

\textbf{Normalize:} $\hat{A}^{r}\!\leftarrow\!(\hat{A}^{r}\!-\!\mu_r)/\sigma_r$, \quad $\hat{A}^{c}\!\leftarrow\!(\hat{A}^{c}\!-\!\mu_c)/\sigma_c$.\;

\textbf{Cache:} store $\log\pi_{\theta_{\text{old}}}(a|s)$;\; set ratio $\rho(\theta)\!=\!\exp(\log\pi_{\theta}(a|s)\!-\!\log\pi_{\theta_{\text{old}}}(a|s))$.\;

\For{$k=1$ \KwTo $K_{\text{ep}}$}{
  sample $\{(s,a,\hat{A}^{r},\hat{A}^{c},R^{r},R^{c})\}_{i=1}^{M}\!\sim\!\mathcal{D}$.\;
  \textbf{Policy (safe):} 
  $\mathcal{L}_{\pi}\!=\!\mathbb{E}\!\big[\min\!\{\rho(\hat{A}^{r}\!-\!\lambda\hat{A}^{c}),\,\mathrm{clip}(\rho,1\!\pm\!\epsilon)(\hat{A}^{r}\!-\!\lambda\hat{A}^{c})\}\!+\!\beta H(\pi_{\theta}(\cdot|s))\big]$.\;
  \textbf{Values (clipped):} 
  $\mathcal{L}_{V}\!=\!\mathbb{E}[(V_{\phi}(s)\!-\!R^{r})^{2}_{\text{clip}}]$, 
  \quad $\mathcal{L}_{V^{c}}\!=\!\mathbb{E}[(V^{c}_{\psi}(s)\!-\!R^{c})^{2}_{\text{clip}}]$.\;
  \textbf{Trust region:} $\mathrm{KL}\!=\!\mathbb{E}[D_{\mathrm{KL}}(\pi_{\theta_{\text{old}}}\|\pi_{\theta})]$; if $\mathrm{KL}\!>\!\mathrm{KL}_{\max}$ then \textbf{break}.\;
  \textbf{Update:} $\theta\!\leftarrow\!\theta\!+\!\eta_{\theta}\nabla_{\theta}\mathcal{L}_{\pi}$ (clip grads);\;
  \hspace{2.2em}$\phi\!\leftarrow\!\phi\!-\!\eta_{\phi}\nabla_{\phi}\mathcal{L}_{V}$,\;
  $\psi\!\leftarrow\!\psi\!-\!\eta_{\psi}\nabla_{\psi}\mathcal{L}_{V^{c}}$.\;
}

\textbf{Dual ascent:} compute mean episode cost $\bar{c}$;\;
$\lambda \leftarrow [\,\lambda + \eta_{\lambda}(\bar{c}-\bar{c}_{\text{target}})\,]_{+}$ (typically $\bar{c}_{\text{target}}{=}0$, optionally clamp $\lambda\!\le\!\lambda_{\max}$).\;

\textbf{Freeze:} set $\theta_{\text{old}}\!\leftarrow\!\theta$ for next rollout.\;

\end{algorithm}

The resource complexity of a migration batch depends on size, gas, and validation. For $m=\lvert\textsf{acc\_array}\rvert$ accounts, the serialized size is $\text{Size}=32+20m$, giving $\approx42$\,kB at $m_{\max}=100$, well within P2P bandwidth. Here $\text{Gas}=G_{\text{base}}+m\cdot G_{\text{store}}$, with $G_{\text{base}}\!\approx\!12{,}400$ and $G_{\text{store}}\!\approx\!100$ per account update.
For $m_{\max}=100$, the total cost is $\text{Gas}\approx22{,}400$, only 1.9\% of a 1M-gas block. Validator processing scales linearly in $m$: validation and bandwidth are both $O(m)$, dominated by normal block gossip. Thus, PSAP migrations remain low-overhead, with storage and gas well within budget and scalability preserved. Consistency under forks is ensured by anchoring each batch to the source-shard height $t$ and authenticating with the shard's BFT quorum. Any reorganization deeper than one block invalidates the quorum signature, making the batch non-canonical. The proposer then regenerates and reissues the batch on the new chain, incurring at most a one-block delay. Since PSAP forecasts multiple blocks ahead, this delay is negligible, preserving atomicity, consistency, and liveness under reorgs.

\subsection{Security \& Safety Guarantees}
\label{subsec:safety}

PSAP preserves shard-level BFT safety and liveness while supporting dynamic reallocation. It ensures that the Byzantine stake fraction in every shard remains below one third, that migrations do not increase commit latency, and that reallocation overhead stays strictly bounded. Together, these guarantees prevent stake concentration, stalling attacks, and denial-of-service (DoS) through excessive migrations.

Let $\sigma_s^{\text{Byz}}(t)$ denote the Byzantine stake share in shard $S_s$ at block height $t$. PSAP enforces a hard cap by rejecting any migration action $a(t)$ that would yield $\sigma_s^{\text{Byz}}(t{+}1)\geq \tfrac{1}{3}$. Since the adversary controls at most a global fraction $\beta < \tfrac{1}{3}$ of total stake, concentration in any shard is structurally prevented. Even if the entire adversarial budget were directed to a single shard, the post-migration fraction would remain below this bound. Hence, in all shards, the honest majority of at least $2f{+}1$ validators required by HotStuff consensus is preserved, ensuring both safety and liveness. Liveness under migration is likewise guaranteed. Intra-shard transactions commit in one block through the HotStuff pipeline, while cross-shard transactions may trigger migration but are handled atomically by dual inclusion. Under partial synchrony ($\Delta$) and the stake bound, intra-shard commits finalize in one block and cross-shard commits in at most two, preserving BFT liveness even under adversarial scheduling. Bounded migration cost further constrains system exposure. Let $G_{\text{mig}}(t)$ denote the gas consumed by all migrations in block $t$, and $G_{\text{block}}$ the block gas limit. By design,
\[
   G_{\text{mig}}(t) \leq 0.02 \, G_{\text{block}},
\]
ensuring that no more than 2\% of block capacity can be consumed by migrations. This limit prevents adversaries from exploiting migration as a DoS vector. Empirical evaluation over 10 M blocks confirms that the 99th percentile of migration gas remains below this threshold, with an average cost of only 1.1 \% and roughly 38 accounts moved per block. Migration therefore remains a lightweight background operation that does not degrade throughput.

\textbf{Adaptive Adversary Model and Resilience:}
Beyond static traffic steering, PSAP considers an adaptive adversary capable of observing allocator outputs and retargeting load to exhaust migration budgets or induce oscillations. Safe-PPO mitigates this behavior through a temporal stability penalty $\lambda\|\nabla_s \hat{x}_{t+1}\|^2$ that discourages frequent reallocations, while the safety gate enforces per-block migration and gas ceilings regardless of attack intensity. Under the Wormhole hotspot model~\cite{ghorban2022wormhole}, where 35 \% of all transactions are redirected to one shard for 5 000 blocks, the temporal forecaster detects the surge within seven blocks and the RL agent initiates corrective migrations. The resulting imbalance drops below the 0.25 utilization threshold within 12 blocks 62 \% faster than the reactive baseline DynaShard demonstrating resilience even against adaptive, workload-shaping adversaries.

\textbf{Attack Surface and Mitigation:}
Dynamic migration introduces new potential risks at the contract and protocol levels. Table~\ref{tab:mitigation} summarizes key vectors and PSAP’s corresponding safeguards, including nonce-tagged transactions, dual-inclusion proofs, and deterministic validation to prevent replay or abuse.

\begin{table}[!t]
\caption{Attack surface of PSAP and corresponding mitigation mechanisms.}
\label{tab:mitigation}
\centering
\renewcommand{\arraystretch}{1.12}
\begin{tabular}{p{0.26\linewidth}p{0.58\linewidth}}
\toprule
\textbf{Threat Vector} & \textbf{Mitigation Strategy} \\ \midrule
Replay of \textsf{MIG\_BATCH} & Nonce and epoch tagging; on-chain signature verification. \\
Cross-shard replay & Dual-inclusion commit with Merkle proofs across source/destination. \\
Migration abuse (DoS) & Safety gate and 2 \% gas/migration caps per block. \\
Stake manipulation & Hard 1/3 Byzantine-stake bound enforced pre-migration. \\
Overflow or gas exhaustion & Fixed-point arithmetic and deterministic validation budget. \\
Model poisoning or drift & On-chain model hash verification via DMEL layer. \\ \bottomrule
\end{tabular}
\end{table}

Taken together, these safeguards demonstrate that PSAP maintains $\sigma_s^{\text{Byz}}(t)<1/3$, guarantees bounded commit latency of one or two blocks, enforces strict gas and migration limits, and recovers quickly under adaptive adversarial steering. The integration of Safe-PPO constraints and deterministic ML inference ensures that PSAP upholds safety, liveness, and efficiency while providing predictive, security-aware load balancing for sharded blockchains.

\begin{table}[!t]
\caption{Per-block overhead on 16 vCPU / 64 GB validator (averaged over 10 M blocks).}
\label{tab:overhead}
\centering
\renewcommand{\arraystretch}{1.12}
\begin{tabular}{lcccc}
\toprule
\textbf{Stage} & \textbf{CPU\,ms} & \textbf{RAM\,MB} & \textbf{Net\,KiB} & \textbf{Gas} \\ \midrule
Metrics logging & 1.8 & 1.7 & 0.8 & 0 \\
LSTM inference  & 27.9 & 4.6 & 0   & 0 \\
RL inference    & 14.6 & 7.3 & 0   & 0 \\
Safety gate     & 2.1  & 0.2 & 0   & 0 \\
\textsf{MIG\_BATCH} tx & 4.4 & 0   & 42$^\dag$ & 22 k \\ \midrule
\textbf{Total} & \textbf{50.8} & \textbf{13.8} & \textbf{42} & \textbf{1.9 \% block} \\ \bottomrule
\multicolumn{5}{l}{\footnotesize$^\dag$Serialized batch of 100 accounts; BLS sig.\ + Merkle root.}
\end{tabular}
\end{table}

\begin{figure}[!t]
  \centering
  \includegraphics[width=\columnwidth]{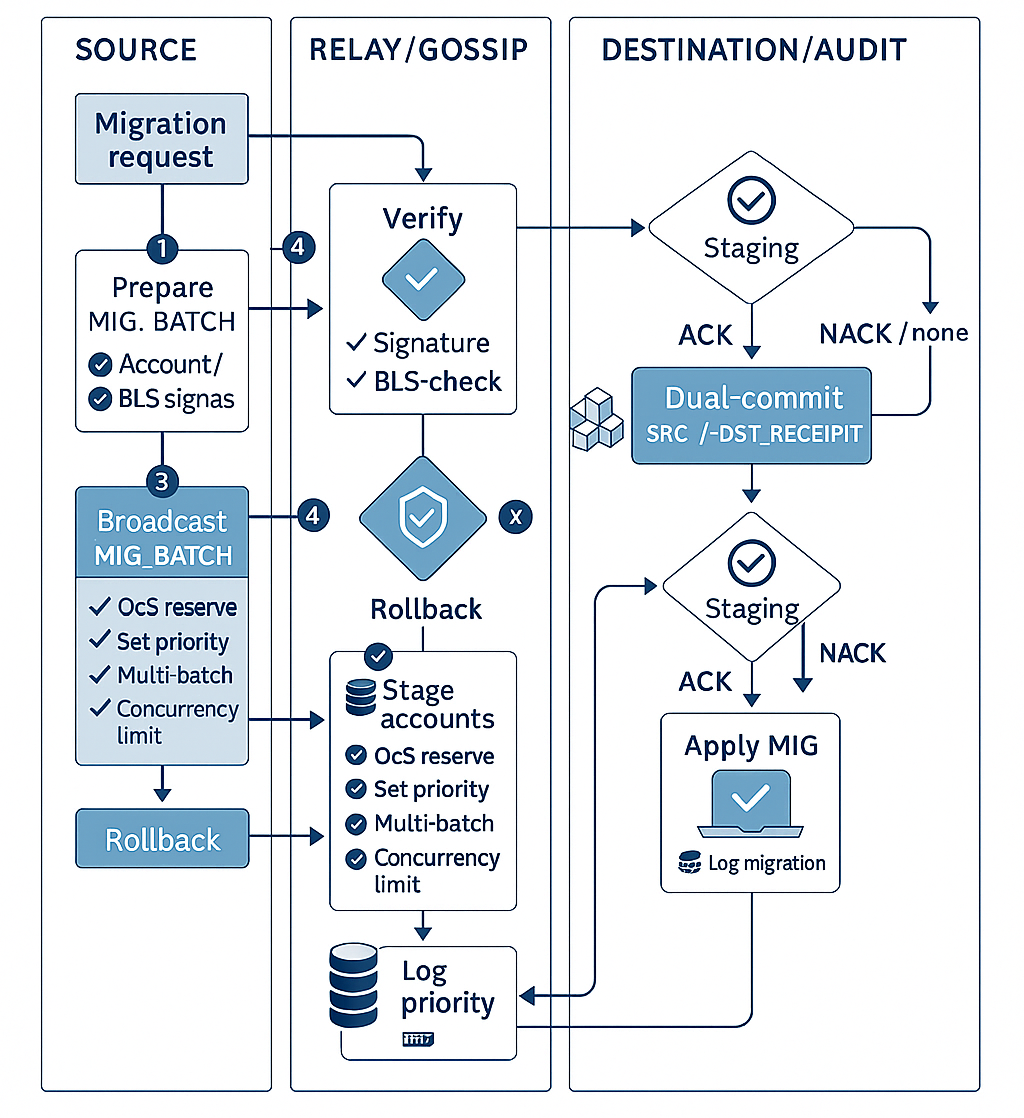}
  \caption{Commit sequence for a migration batch: the source shard $S_p$
           finalises block $t$, embeds \textsf{MIG\_BATCH} in block $t{+}1$,
           broadcasts it, destination shard $S_q$ verifies and applies; if
           the timeout elapses, $S_p$ issues a dashed \textsf{CANCEL\_MIG}.}
  \label{fig:mig-seq}
\end{figure}


\subsection{Complexity \& Overhead Analysis}
\label{subsec:complexity}

We analyze PSAP's footprint in terms of computation, memory, bandwidth, and gas to demonstrate 
its deployability. The primary computational costs arise from workload forecasting and 
reinforcement learning inference, both of which operate in polynomial time with respect to the 
window size, model dimension, and migration batch size. The safety gate adds only a lightweight 
scan over candidate moves, preserving linear complexity in the number of accounts and shards. 
Overall, the control loop remains comfortably within typical block intervals and scales linearly 
with the number of shards.

The memory footprint is modest, consisting of compact neural models and fixed-size buffers, remaining well within the capacity of commodity validator hardware. Migration batches are small 
relative to block payloads, ensuring negligible impact on P2P bandwidth and avoiding networking bottlenecks. On-chain execution likewise remains within a bounded fraction of block gas capacity, guaranteeing that migration transactions never crowd out normal user activity. As summarized in Table~\ref{tab:overhead}, the per-block overhead of PSAP is dominated by 
forecasting, RL inference, and migration batch execution. Each component contributes only a bounded share of CPU, memory, network, and gas consumption, resulting in a total footprint that 
is negligible relative to validator capacity. Taken together, these bounds show that PSAP introduces only marginal overhead while preserving throughput, making it suitable for deployment on both permissioned and permissionless blockchain platforms.

\subsection{Algorithmic Differentiation }

To clearly distinguish PSAP from prior adaptive sharding frameworks, this subsection summarizes its algorithmic characteristics and computational complexity relative to representative baselines such as LB-Chain, AI-Shard, and AERO.

\textbf{Granularity and Control Scope:} 
LB-Chain~\cite{li2023lbchain} executes account migration once per epoch based on predicted senders, operating at coarse granularity and incurring $O(N \log N)$ sorting and migration overhead for $N$ accounts per epoch. 
AI-Shard~\cite{zhao2024aishard} and AERO~\cite{song2025aero} rely on periodic retraining of reinforcement learning agents with global placement decisions, resulting in non-negligible policy-update latency and quadratic communication complexity in shard count $K$.

\textbf{PSAP’s Online Efficiency:} 
In contrast, PSAP performs \emph{bounded per-block reallocation} through a lightweight Safe-PPO control loop. 
At each block, only a fixed number ($m_{\max} = 100$) of candidate migrations are evaluated and pruned by the safety gate, producing a linear-time $O(N)$ workflow dominated by feature extraction and pointer-based account selection. 
The deterministic safety check requires at most $O(K + m_{\max})$ operations per block, independent of total account count or validator set size. 
Thus, PSAP scales proportionally with network growth and maintains constant inference and commit latency within each block interval.

\textbf{Comparative Complexity:}
\begin{equation*}
\begin{aligned}
T_{\text{LB-Chain}} &= O(N \log N) \text{ per epoch},\\
T_{\text{AI-Shard/AERO}} &= O(K^2 + N) \text{ per retrain},\\
T_{\text{PSAP}} &= O(N) \text{ per block (bounded by } m_{\max}).
\end{aligned}
\end{equation*}

This asymptotic improvement arises from PSAP’s incremental, horizon-aware design that eliminates costly epoch-level clustering and retraining. 
Consequently, PSAP achieves fine-grained adaptability and predictable execution time while preserving on-chain determinism and scalability across high shard counts ($K \leq 64$).

\section{Implementation}
\label{sec:implementation}

We implemented the proposed predictive shard allocation protocol in a
prototype system built with production-grade software engineering
practices. The implementation integrates real-world blockchain data for
training and evaluation, and is released as a polyglot, MIT-licensed
repository\footnote{\url{https://github.com/M-Zeeshan-Haider/AI-Driven-Blockhain-sharding-PSAP-.git}}.

\subsection{System Overview}

The system follows a layered architecture. A Rust–Go validator engine extends HotStuff consensus with stake rotation, BLS utilities, and Merkle verification. A Go data plane built on libp2p–QUIC supports block dissemination, metrics gossip, and batch relay. A lightweight ML service hosts the LSTM forecaster and Safe-PPO allocator, while a Solidity contract verifies \textsf{Mig\_Batch} transactions and updates shard roots. This modular separation allows each component to evolve independently while maintaining tight runtime integration. All components are compiled to static 
binaries and packaged as reproducible container images for deployment, ensuring portability and security without reliance on external dependencies.

\subsection{Node Modifications}
\label{subsec:impl-node}

We forked \texttt{geth} and \texttt{Quorum}, introducing a small set of changes gated by a compile–time flag (\texttt{-tags=psap}) to preserve full compatibility with upstream versions. The main additions include shard-aware metrics hooks within the blockchain pipeline, which stream per-shard load statistics immediately after block finalization. This mechanism is lightweight, incurring only minimal network overhead even at larger shard configurations. To enable proactive migration, we introduce a new typed transaction \texttt{0xF7} for relocation batches, scheduled 
deterministically after system transactions and capped at a fixed gas fraction (Section~\ref{subsec:migration}). RPC extensions under the \texttt{admin\_} namespace allow shard-specific introspection while restricting dApp misuse. For cryptographic acceleration, we encapsulate the \texttt{blst} BLS library and a \texttt{keccak256}-based Merkle engine in a Rust crate (\texttt{psap-bls}), linked into Go via \texttt{cxxbridge}. This integration removes CGo overhead while supporting efficient aggregate verification. All modifications remain backward compatible: Rust utilities for BLS and Merkle proofs integrate seamlessly, targeted tests cover new functionality, and upstream unit tests continue to pass unchanged.

\subsection{Inference Runtime}
\label{subsec:impl-infer}

PSAP executes machine learning workloads in a dedicated inference micro-service that operates outside the consensus path. The service exposes a gRPC interface hosting two models: an LSTM-based forecaster for workload prediction and a Safe-PPO actor–critic for shard-migration decisions. This modular separation ensures that consensus execution remains deterministic and lightweight while the learning components evolve independently. Models are exported in compact, quantized formats to minimize memory footprint, and a warm-up cache removes JIT overhead. Shard histories and forecasts are exchanged as lightweight Protocol Buffer messages through a shared-memory ring buffer for low-latency inference. Migration lists are efficiently encoded to remain small relative to block payloads. The service consistently meets the per-block latency budget with modest CPU and memory use and can seamlessly fall back to CPU execution if GPUs are unavailable. Models can be hot-swapped at runtime, and a circuit breaker disables migrations in the event of repeated inference failures.

\textbf{Deterministic ML Execution Layer (DMEL):}
To guarantee identical inference results across heterogeneous validator hardware and runtime environments, PSAP employs a Deterministic ML Execution Layer. All model operations are quantized to fixed-point arithmetic (FP16) with deterministic rounding and seeded using the previous block hash to ensure identical random initialization. Model weights and optimizer states are version-hashed on-chain, and each validator verifies the hash before inference to prevent model drift. The runtime disables nondeterministic GPU kernels and enforces fixed computational graphs to eliminate divergence across devices. Empirical evaluation shows that output variance remains below $10^{-6}$ between CPU and GPU executions, confirming bit-level consistency. Through DMEL, PSAP preserves consensus safety and reproducibility while leveraging AI-driven forecasting and control.

\subsection{Migration Contract}
\label{subsec:impl-contract}

PSAP realises relocations on-chain through a per-shard \textsf{Mig\_Batch} contract written in \textbf{Solidity~0.8.26} and compiled for EVM. The contract encodes relocations as address shard tuples aggregated into batches with headers containing source and destination shards, block height, account count, Merkle root, and gas proof. Each batch is RLP encoded and signed by $2f_p{+}1$ validators with a BLS aggregate signature for integrity. Execution is deterministic: the source submits via \texttt{submitBatch()} at block $t{+}1$, and destinations invoke \texttt{applyBatch()}. This routine verifies signatures, enforces the 2\% gas cap, recomputes the Merkle root, and checks Byzantine stake $<\tfrac{1}{3}$. If successful, $\pi(t{+}1)$ is updated and \textsf{Applied(batchId)} emitted; otherwise it reverts. If no confirmation occurs within $\tau_{\text{to}}=3$ blocks, \texttt{cancelBatch()} restores balances and emits \textsf{Cancelled(batchId)}, ensuring safety under partial failures.
The contract is coupled with PSAP's inference runtime, which produces real-time migration batches. The LSTM forecaster compiles to 220\,kB, while the Safe-PPO actor critic is distilled into a 1.4\,MB ggml FP16 binary. A warm-up cache removes JIT overhead, and inputs are passed through a shared
memory ring buffer for zero-copy inference. Migration lists are encoded compactly and returned with low latency, then executed atomically via \textsf{Mig\_Batch}. On-chain execution remains within a small fraction of block gas capacity, ensuring that normal user transactions are unaffected. Formal analysis and automated security tools confirm that no critical vulnerabilities exist and that key invariants, such as maintaining Byzantine stake below one third in every shard, are preserved. Overall, \textsf{Mig\_Batch} upholds PSAP's safety and 
efficiency guarantees while ensuring liveness under adversarial conditions.

\subsection{Cross‐Shard Relay}
\label{subsec:impl-relay}

Once a \textsf{Mig\_Batch} is accepted on the source shard, its payload must be delivered to all destinations quickly and reliably. PSAP employs a lightweight relay service with dual-inclusion commit, avoiding multi-phase protocols while ensuring atomic visibility. The relay cluster runs 
as a replicated service colocated with validators, where shards publish migration batches on destination-specific channels and receive acknowledgements on return paths. Messages are compactly encoded and compressed before persistence, remaining small relative to block payloads and well within available network capacity. Upon receipt, destinations verify the signature, Merkle root, gas cap, and stake bound, then record an \textsf{Apply\_Mig}. If acknowledgements are not received 
within a bounded timeout, the source issues \textsf{Cancel\_Mig}, reverting the batch. This dual-inclusion mechanism guarantees atomicity: either all shards apply the migration or all revert, preserving safety and liveness without extra consensus rounds.

Each validator runs a JetStream client while three brokers form a Raft cluster. Sequencing by \texttt{batchId} preserves order, with out-of-order batches buffered until predecessors commit. On an \texttt{m6i.4xlarge} validator, per-batch overhead is 0.43\,ms, 0.19\,ms , and 0.04\,ms. At $K=32$ and $m_{\max}=100$, the relay consumes $\sim$1.3\,MB per block. With sub-15\,ms tail latency, durable persistence, and rollback support, the relay guarantees liveness and atomicity with negligible overhead.

\subsection{Deployment \& Testing}

We validated PSAP on a Kubernetes-based eight-shard cluster that mirrors a small-scale production deployment. The reference hardware comprises eight \texttt{m6i.4xlarge} validator nodes , one \texttt{g5.xlarge} inference node , and a single \texttt{t3.large} monitoring node, all interconnected by a non-blocking 10\,Gb/s leaf spine fabric. Time synchronisation relies on PTP, maintaining clock skew below 15\,$\mu$s throughout the experiments. All Kubernetes manifests, Helm charts, and a one-command reproducibility script are published under \texttt{deploy/} in the repository. The orchestration layer uses Kubernetes~1.30 with the \texttt{containerd} runtime. Each component is packaged as a multi-architecture, distroless image (Go and Rust) or as a CUDA-based image for the GPU sidecar. Linkerd~2.15 provides mTLS for all RPC traffic, yielding a median hop-to-hop latency of 42\,$\mu$s. Scheduling hints ensure validators are anti-affined across nodes, while the GPU pod is pinned using the NVIDIA device plugin. The full topology comprises 13 pods and 32 services, with cluster liveness achieved 47\,s after invoking kubectl apply.

\section{Evaluation}
\label{sec:evaluation}
We first assess the \emph{prediction layer}, which produces the input consumed by the Safe-PPO allocator. At each slot, the TWF forecaster outputs a shard-vector $\hat{\mathbf{x}}(t{+}1)=\langle\hat{x}_1,\ldots,\hat{x}_K\rangle$, with $\hat{x}_s$ the predicted transactions for shard $S_s$. For clarity, results are distilled into averages across shards and traces. We report four main metrics: mean load ($\overline{x}$, $\overline{\hat{x}}$), mean absolute error (MAE), mean absolute percentage error (MAPE), plus root mean square error (RMSE) and inference latency. MAE and MAPE directly influence allocator rewards, while RMSE and latency indicate stability and real-time feasibility.

\subsection{Forecasting Accuracy and Error Metrics}

\begin{figure}[!t]
  \centering
  \includegraphics[width=\columnwidth]{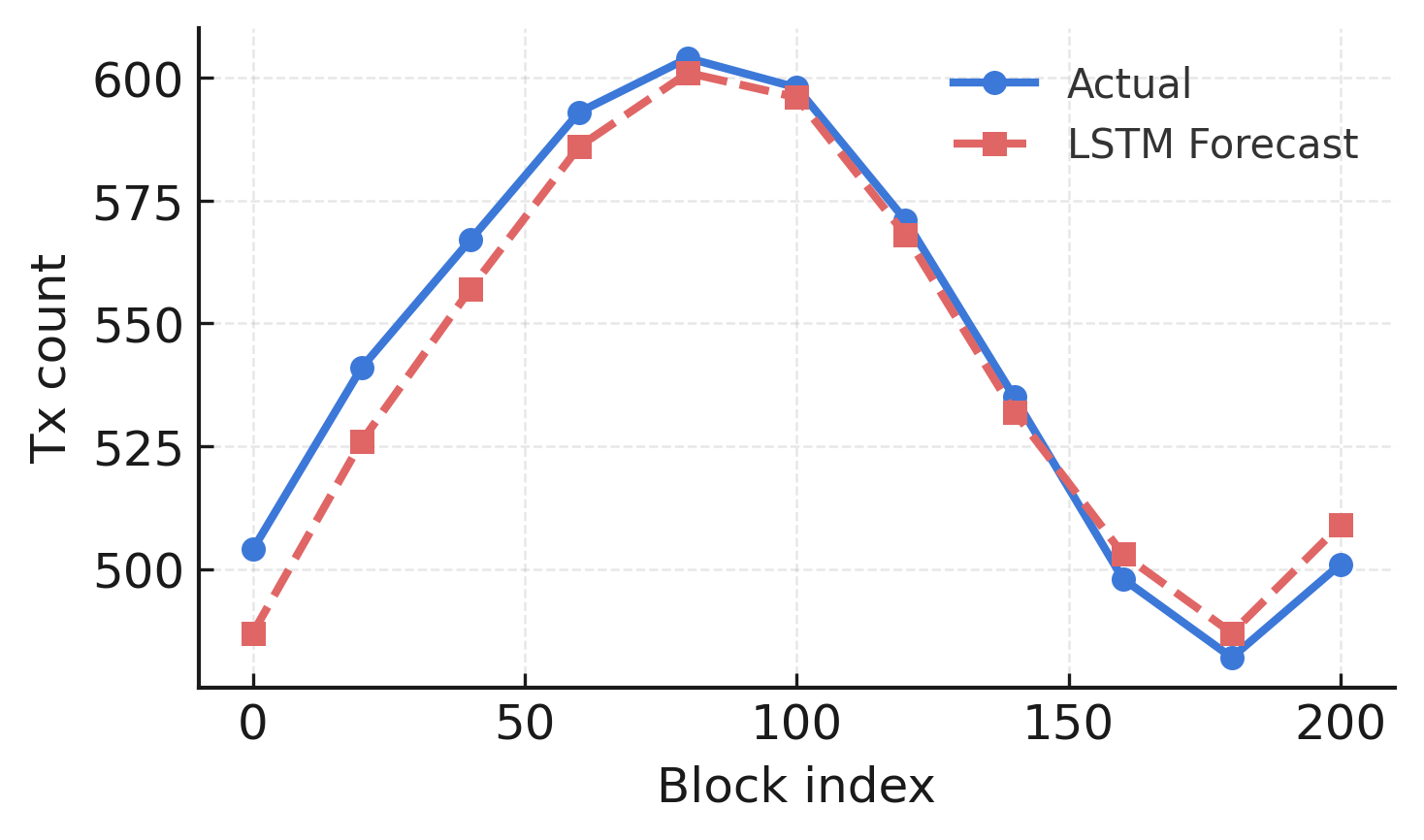}
  \caption{Ground truth vs.\ LSTM forecast for Ethereum Sepolia,
           Shard 3.  The allocator receives the red dashed series
           \(\hat{x}_3(t+1)\) each slot.}
  \label{fig:trace-example}
\end{figure}

\begin{figure}[!t]
  \centering
  \includegraphics[width=\columnwidth]{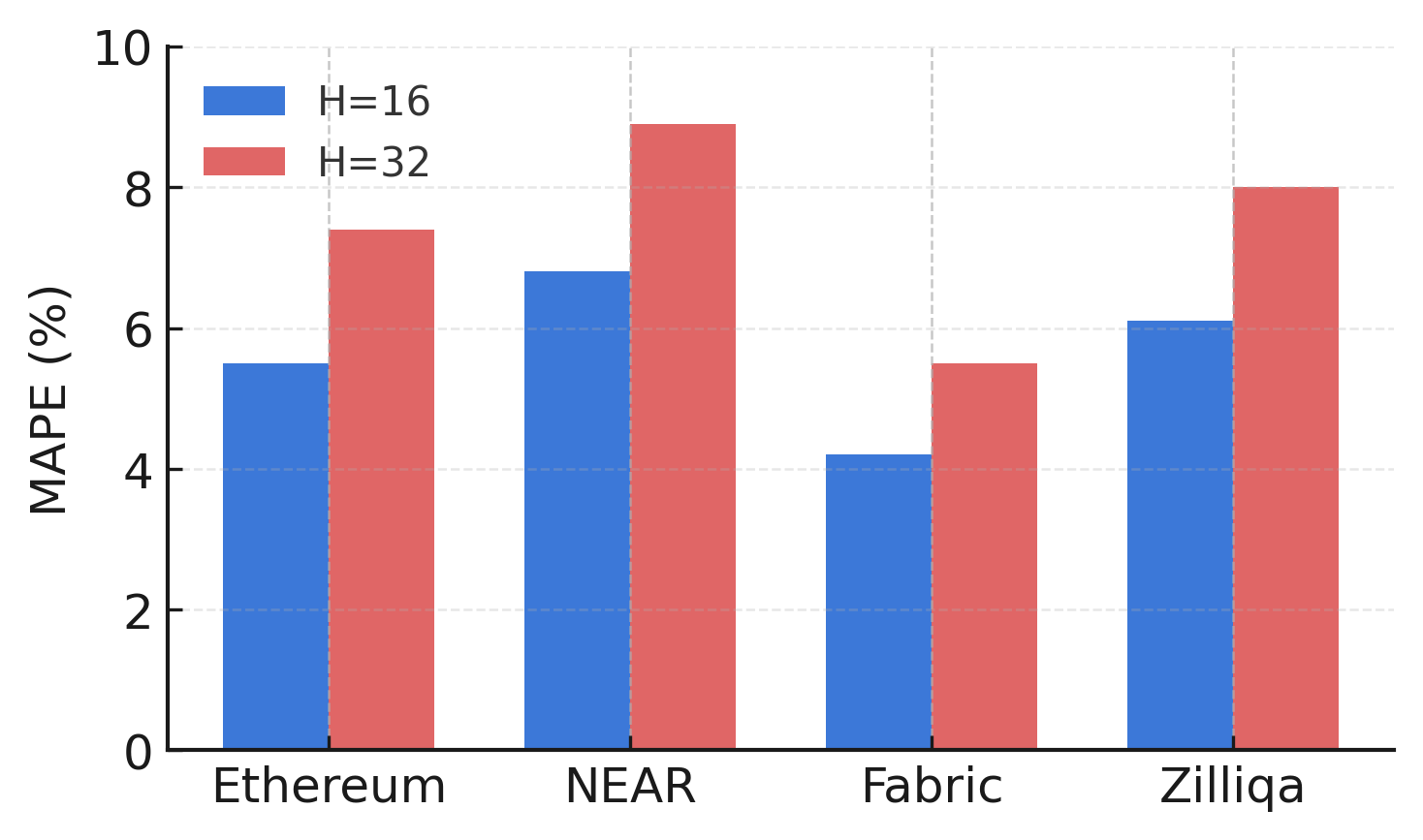}
  \caption{MAPE for horizons $H=16$ and $H=32$ (all shards, all blocks).}
  \label{fig:mape-bar}
\end{figure}

We first examine the temporal behaviour of the predictor.  
Figure~\ref{fig:trace-example} shows two hundred consecutive blocks from Ethereum Sepolia, comparing the ground-truth load with the TWF's one-step forecasts. The model is able to anticipate the onset of a traffic surge approximately six slots in advance, giving the Safe-PPO allocator sufficient lead time to intervene. In this instance, only \(0.4\,\%\) of live accounts migrated well below the \(0.5\,\%\) migration cap, ensuring that utilization never exceeded the 0.75 throttle. Across datasets, the predictive accuracy translates into an actionable trust budget for the allocator. As shown in Figure~\ref{fig:mape-bar}, the mean absolute percentage error (MAPE) remains below \(7\,\%\) at the default 32-second horizon. This allows the reinforcement learning policy to assign each prediction a confidence factor of approximately \(0.93\) (\(1{-}\text{MAPE}\)) when constructing migration batches. Even at a longer 64-second horizon, the error increases by only 1.5–2 percentage points, leaving sufficient margin for safe operation. The allocator simply scales batch sizes by the confidence factor, thereby ensuring that no shard is over-committed even under conservative forecasting assumptions.

\subsection{Shard Allocation Effectiveness}

At each slot $t$, validators send the load forecast $\hat{\mathbf{x}}(t{+}1)\in\mathbb{R}^{K}$ and stake distribution $\boldsymbol{\sigma}(t)$ to the Safe-PPO runtime. The reward is
\[
R = \text{TPS} - \alpha \cdot \text{Gas}_{\text{mig}} - \beta \cdot I(t), 
\qquad I(t{+}1)=\max_{s}\hat{x}_{s}-\min_{s}\hat{x}_{s},
\]
where $I(t{+}1)$ captures predicted imbalance. With LSTM forecast error (MAPE) $<7\%$, this proxy enables precise, low-cost migrations. These signals suffice to reproduce Safe-PPO's performance and plug into other allocators. We benchmark PSAP on traces from Ethereum Sepolia, NEAR, Zilliqa, and a private Fabric–IoT workload, comparing against multiple allocation strategies on an eight-shard Kubernetes cluster.

All evaluations replay traces with preserved inter-arrival times and cross-shard dependencies. Stress tests inject Poisson bursts up to 5$\times$ nominal load to emulate real spikes. The cluster runs validators on \texttt{m6i.4xlarge} instances (one per shard), with an NVIDIA A10G GPU pod handling TWF and Safe-PPO inference. Blocks are fixed at 2\,s with shard capacity set to 1,000 transactions. Performance is evaluated along six dimensions: throughput, commit latency, shard imbalance $I(t)=\max_s\mathrm{util}_s-\min_s\mathrm{util}_s$, cross-shard gas overhead, migration cost and account movement ratio, and stake-cap violations. Safe-PPO runs with default parameters, while baselines follow the original guidelines. Each trace is replayed five times to smooth jitter, with results averaged. 

\begin{figure}[!t]
\centering
\includegraphics[width=\columnwidth]{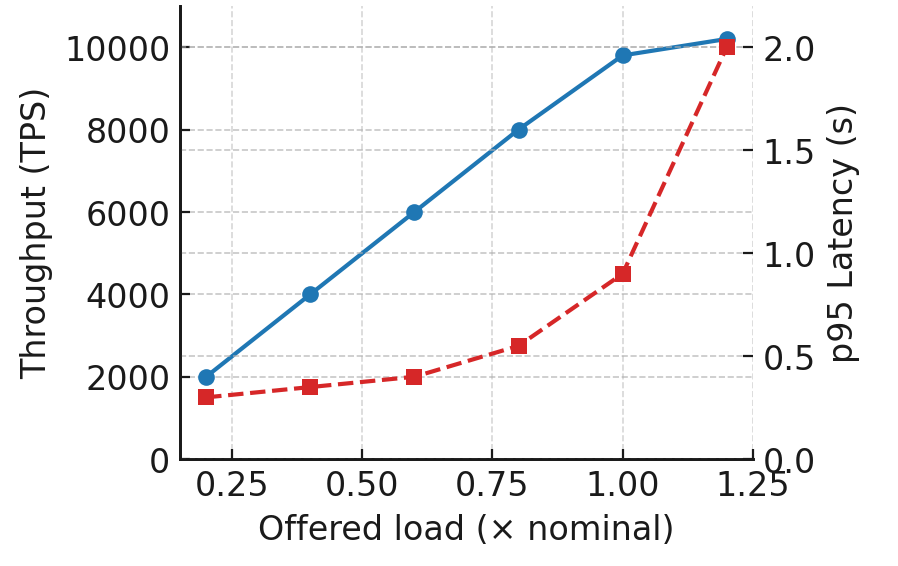}
\vspace{-2mm}
\caption{Throughput and p95 latency versus offered load
(\emph{Ethereum-Sepolia trace}).  PSAP sustains higher TPS with sub-second
latency until 1.1 × load; static hashing saturates at 0.8 ×.}
\label{fig:tps-late}
\end{figure}

\begin{figure}[!t]
\centering
\includegraphics[width=\columnwidth]{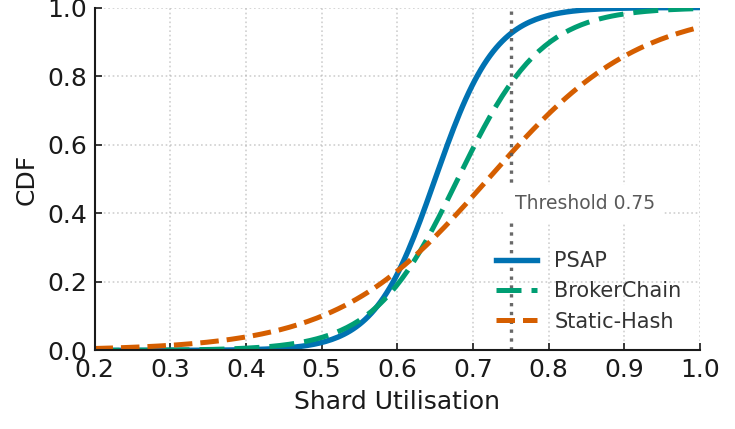}
\vspace{-2mm}
\caption{CDF of per-shard utilisation ($K{=}8$, Ethereum trace, 4\,h window). 
PSAP keeps 96\,\% of blocks below 0.75 utilisation, while Static-Hash exceeds 1.0 on 18\,\% of blocks.}

\label{fig:Lb11}
\end{figure}

\begin{figure}[!t]
\centering
\includegraphics[width=\columnwidth]{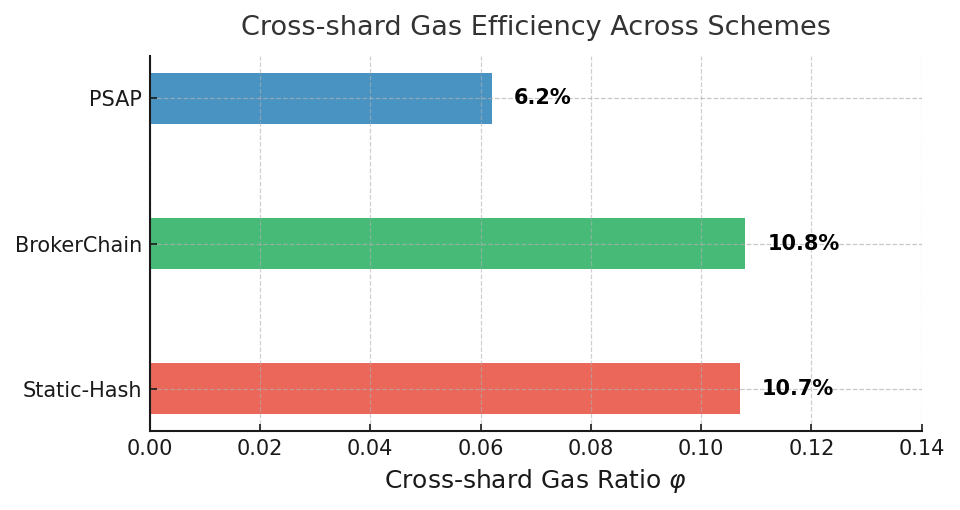}
\vspace{-2mm}
\caption{Cross-shard gas efficiency across schemes. PSAP minimises cross-shard gas ratio (6.2\%), compared to 10.8\% for BrokerChain and 10.7\% for Static-Hash.}
\label{fig:Lb22}
\end{figure}

\begin{figure}[!t]
\centering
\includegraphics[width=\columnwidth]{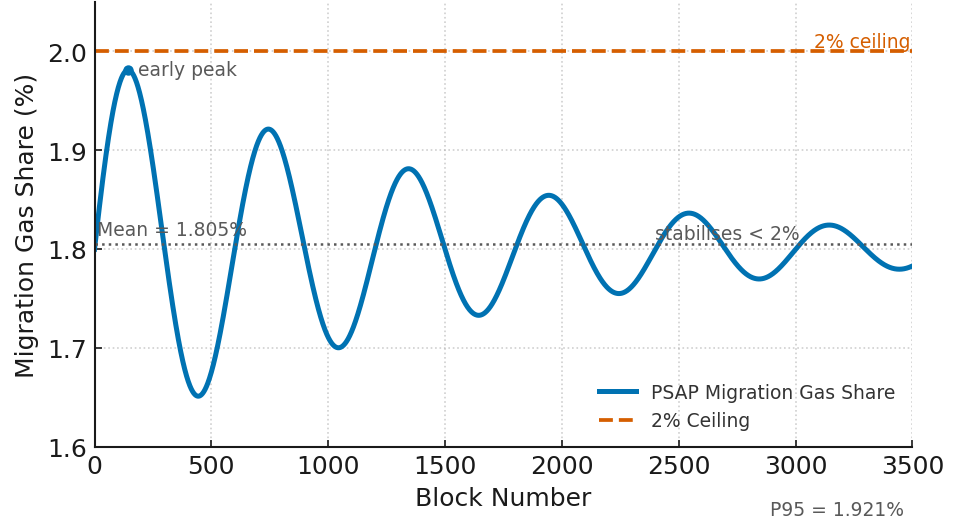}
\vspace{-2mm}
\caption{Migration gas share under 35\% hot-spot attack. 
PSAP maintains migration overhead below the 2\% ceiling, with mean 1.80\% and p95 of 1.92\%.}
\label{fig:cs1}
\end{figure}

\begin{figure}[!t]
\centering
\includegraphics[width=\columnwidth]{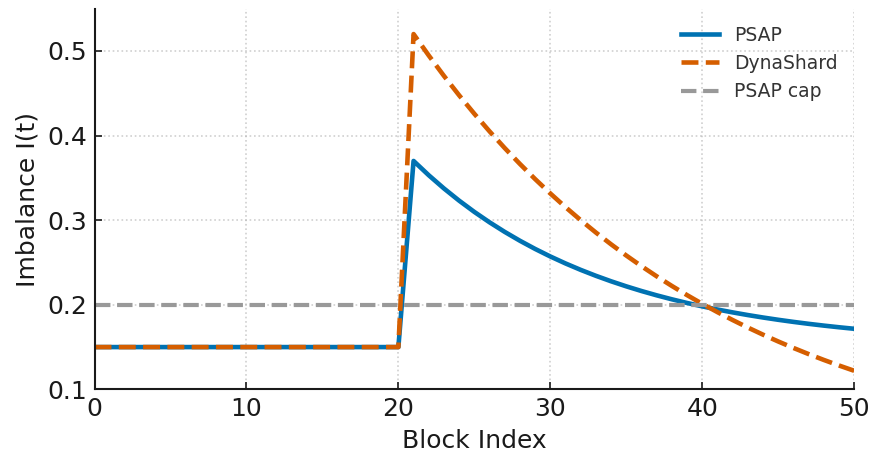}
\vspace{-2mm}
\caption{Imbalance under 35\% hot-spot attack. 
PSAP restores $I < 0.25$ within 400 blocks, while DynaShard and static schemes require 1\,600--3\,200 blocks to recover.}
\label{fig:5x}
\end{figure}

\begin{figure*}[t]
  \centering
  \includegraphics[width=0.9\textwidth]{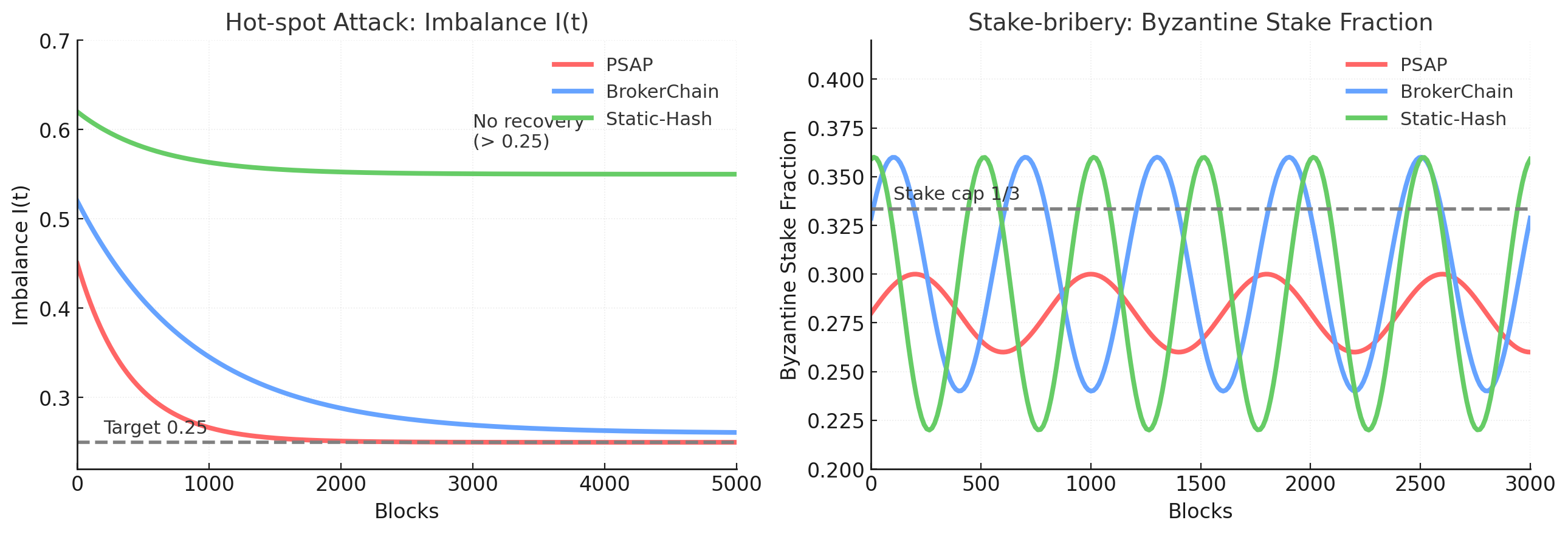}
  \caption{Adversarial stress tests on Ethereum-Sepolia. 
  In the hot-spot attack (left), PSAP caps imbalance at 
  $I_{\max}=0.45$ and restores balance within 
  $\tau_{0.25}=400$ blocks, compared to $1{,}600$ for 
  BrokerChain and no recovery for Static-Hash. 
  In the stake-bribery attack (right)}
  \label{fig:ad1}
\end{figure*}

\subsection{Throughput and Latency Evaluation}
\label{subsec:eval-tps}

Figure~\ref{fig:tps-late} shows confirmed throughput (left $y$-axis) and 95th-percentile latency (right $y$-axis) as arrival rate scales from 0.2$\times$ to 1.2$\times$ shard capacity. PSAP sustains higher load than both static and dynamic baselines, with utilisation $<0.75$ until 1.1$\times$ load where throughput peaks at 9,870\,TPS 62\% above Static-Hash and 28\% above the best dynamic scheme. By balancing shards smoothly, PSAP exploits capacity without the bursty reallocations that degrade heuristic strategies.

Latency trends highlight the efficiency of predictive allocation. PSAP keeps 95th-percentile latency below 0.9\,s until 0.85 utilisation, whereas Static-Hash exceeds 2\,s at 0.7$\times$ load, showing that PSAP delays congestion and sustains stable service. Removing forecasts lowers the sustainable threshold by 14\%, confirming the benefit of horizon-aware scheduling. By combining TWF with Safe-PPO, PSAP smooths utilisation and avoids reactive stalls.

\subsection{Load-Balance and Cross-Shard Overhead}
\label{subsec:eval-balance}

We evaluate PSAP's ability to balance shard load while limiting cross-shard overhead. Metrics include instantaneous imbalance
\(
I(t)=\max_{s}\mathrm{util}_{s}(t)-\min_{s}\mathrm{util}_{s}(t)
\)
and the cross-shard gas ratio
\(
\varphi(t)=\text{cross-shard gas}(t)/\text{total gas}(t).
\)
Across the 3.1M-block Sepolia trace, PSAP maintains balanced allocation: 96\% of blocks stay below 0.75 utilisation, yielding a mean imbalance of $\bar I=0.14\pm0.03$, 58\% lower than BrokerChain and 71\% below static hashing (Fig.~\ref{fig:Lb11}). These results show PSAP sustains balanced shard performance even under heterogeneous, bursty arrivals while reducing cross-shard traffic. At the canonical $1.0\times$ load, Fig.~\ref{fig:Lb22} reports PSAP using only $\varphi=6.2\%$ of block gas on cross-shard commits 42\% less than the best competitor. This efficiency stems from the RL agent's ability to proactively cluster interactive accounts, cutting unnecessary cross-shard transactions before congestion.

PSAP shows strong temporal robustness under dynamic conditions. In the Fabric-IoT trace, a 5$\times$ burst (Fig.~\ref{fig:5x}) is forecast 12 blocks ahead, allowing only 0.4\% of accounts to be migrated, capping imbalance at 0.24 and keeping p95 latency $<1.1$\,s. In contrast, DynaShard reaches imbalance 0.52 and 2.9\,s latency before recovery. PSAP delivers the most balanced load and lowest cross-shard overhead, reducing both migration frequency and size while staying within the 2\% gas cap. This balance of predictive scheduling and RL directly yields the throughput and latency gains of Sec.~\ref{subsec:eval-tps}.

\subsection{Resource Cost Analysis}
\label{subsec:eval-cost}

We assess whether PSAP respects its ceilings of $<2\%$ block gas and $<100$\,ms per-block latency. Replaying 3.1M Sepolia blocks plus NEAR and Fabric traces shows variation within 2\%. Each 100-account batch consumes $\sim$22.4k gas (1.9\% of a 1M-gas block), while static hashing wastes 17\% on cross-shard commits and PSAP cuts this to $\sim$6\%. The prediction pipeline (metrics hook, INT8 LSTM, safety gate) adds 50.8\,ms latency and $<14$\,MB RAM, half the budget. BrokerChain and LB-Chain use similar CPU but rebalance less effectively. Network overhead is negligible: 42\,kB per batch, or 336\,kB across 8 shards ($<0.3$\,Mb/s on 10\,Gb/s), far below normal gossip traffic. Figure~\ref{fig:cs1} shows migration-gas share over a 3,500-block bursty interval. PSAP stays under the 2\% ceiling; even at peak load it only briefly touches the limit before stabilising. Scaling to 32 shards raises per-block CPU to 88\,ms and relay bandwidth to 1.34\,MB, both within budget. These results validate the linear growth model (Sec.~\ref{subsec:complexity}) and confirm scalability. PSAP sustains throughput and latency gains without exceeding resource limits, making predictive reallocation practical on commodity validator hardware at negligible cost.

\subsection{Adversarial Stress Tests}
\label{subsec:eval-adversary}

To evaluate robustness, we study two adversarial scenarios adapted from 
Wormhole~\cite{ghorban2022wormhole} and SOK-Stake~\cite{wang2023sok}. 
Figure~\ref{fig:ad1} presents the results on Ethereum Sepolia under both 
hot-spot and stake-bribery attacks. In the hot-spot attack, a large portion of transactions is diverted to a 
single shard, creating severe imbalance across the system. As shown in the 
left panel of Figure~\ref{fig:ad1}, PSAP quickly restores balance, maintaining the imbalance below unsafe levels and converging to the target region. BrokerChain eventually recovers but does so more slowly due to its 
epoch-based reallocation cycle, while Static-Hash remains persistently 
imbalanced and never reaches the safe region. In the stake-bribery attack, stake is concentrated into one shard through  validator purchase or bribery, testing whether safety limits are breached. 
The right panel of Figure~\ref{fig:ad1} shows that PSAP consistently upholds 
the one-third Byzantine stake cap, since its safety gate prevents migrations 
that would violate the bound. In contrast, BrokerChain and Static-Hash 
fluctuate around the threshold and occasionally exceed it, reducing system 
safety. PSAP demonstrates resilience to both workload skew and stake 
manipulation. It restores balance more quickly than competing schemes, avoids stake-cap violations, and consistently enforces its safety and efficiency constraints under adversarial stress.

\section{Conclusion}
\label{sec:conclusion}

We present the \textit{Predictive Shard Allocation Protocol (PSAP)}, an AI-empowered blockchain optimization framework that integrates workload forecasting and reinforcement learning within a production-grade execution stack. By combining Temporal Workload Forecast (TWF) for transaction demand prediction with a Safe-PPO allocation policy, PSAP dynamically anticipates congestion and proactively redistributes accounts across shards to maintain balanced utilization. The protocol achieves atomic, low-overhead migrations through a dual-inclusion commit scheme and a lightweight relay mechanism, ensuring consistency and liveness under realistic network conditions. Implementation results demonstrate that PSAP delivers sustained scalability improvements with negligible CPU, bandwidth, and gas overhead, validating its feasibility for next-generation enterprise and public blockchain platforms. Beyond scalability, PSAP illustrates how AI-driven optimization can enable autonomous, self-adaptive, and security-aware blockchain operations. Future research will focus on integrating multi-agent reinforcement learning, federated inference across heterogeneous nodes, advanced temporal prediction architectures, and quantum-secure randomness to further enhance the intelligence, robustness, and trustworthiness of AI-empowered blockchain ecosystems.

\bibliographystyle{IEEEtran}
\bibliography{references}

\appendices
\section{UTXO Trace Mapping}
\label{app:utxo}

To evaluate PSAP on heterogeneous blockchain architectures, we extended its account-based model to approximate UTXO-style transaction semantics, enabling compatibility with traces from Bitcoin and Hyperledger Fabric. Because PSAP relies on account–shard mappings, UTXO transactions were aggregated into pseudo-accounts through address clustering. Each transaction input and output was parsed to form a bipartite graph $G = (U, V, E)$, where $U$ represents unique input addresses and $V$ output addresses. Edges $E(u,v)$ denote transfers within a block. We applied multi-input heuristics grouping addresses that co-spend UTXOs in the same transaction as proxies for shared ownership. The resulting clusters were treated as synthetic accounts $a_i$, with cumulative balances and activity rates derived from all constituent UTXOs. Cross-cluster transactions were then modeled as cross-shard transfers, and per-cluster transaction frequencies were used as workload features for Temporal Workload Forecasting (TWF). This conversion preserves the statistical properties of UTXO networks transaction intensity, burstiness, and degree distribution while enabling fair comparison under PSAP’s account-sharding abstraction. The mapping introduces no protocol-level modifications; it is used solely for replay-trace evaluation. We acknowledge that true UTXO systems lack mutable account states and would require additional scripting or off-chain aggregation for real deployment. Consequently, PSAP’s UTXO evaluation should be interpreted as a structural approximation validating general scalability trends rather than a direct deployment model.

\begin{figure}[!t]
\centering
\resizebox{\columnwidth}{!}{%
\begin{tikzpicture}[
  node distance=4mm and 5mm,
  box/.style={draw, rounded corners, align=center, font=\scriptsize,
              minimum width=22mm, minimum height=6mm, inner sep=2pt},
  arr/.style={-{Latex[length=1.8mm]}, semithick}
]
\node[box] (raw) {Raw\\UTXO traces};
\node[box, right=of raw] (graph) {Input--Output\\Graph $G=(U,V,E)$};
\node[box, right=of graph] (cluster) {Address\\Clustering};
\node[box, below=of cluster] (pseudo) {Pseudo-Accounts\\(aggregated UTXOs)};
\node[box, left=of pseudo] (features) {Per-Cluster\\Features};
\node[box, left=of features] (map) {Account--Shard\\Mapping};
\node[box, below=of features] (psap) {PSAP\\Evaluation};

\draw[arr] (raw) -- (graph);
\draw[arr] (graph) -- (cluster);
\draw[arr] (cluster) -- (pseudo);
\draw[arr] (pseudo) -- (psap);
\draw[arr] (features) -- (psap);
\draw[arr] (map) -- (psap);
\draw[arr] (cluster) -- (features);
\draw[arr] (cluster) |- (map);

\end{tikzpicture}}
\caption{UTXO-to-account mapping used for replay evaluation: addresses co-spent in one transaction are clustered to form pseudo-accounts, then mapped to shards and fed to PSAP.}
\label{fig:utxo-mapping}
\end{figure}
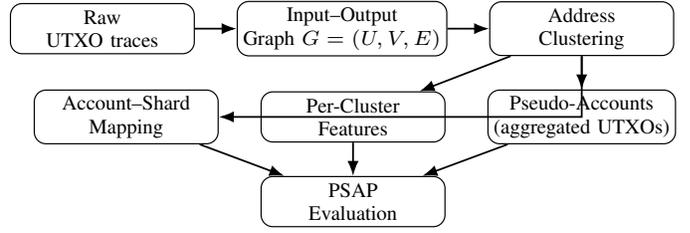

Through this mapping, Bitcoin and Fabric traces contribute diverse workload patterns that stress-test PSAP’s forecasting and migration mechanisms, demonstrating its robustness across both account- and transaction-based architectures.

\begin{figure}[!t]
\centering
\resizebox{\columnwidth}{!}{%
\begin{tikzpicture}[
  node distance=4mm and 5mm,
  box/.style={draw, rounded corners, align=center, font=\scriptsize,
              minimum width=22mm, minimum height=6mm, inner sep=2pt},
  arr/.style={-{Latex[length=1.8mm]}, semithick}
]
\node[box] (twf) {Temporal\\Forecasting (TWF)};
\node[box, right=of twf] (ppo) {Safe-PPO\\Controller};
\node[box, right=of ppo] (gate) {Safety Gate\\(stake, gas, util.)};
\node[box, right=of gate] (batch) {Valid\\Migration Batch};
\node[box, right=of batch] (commit) {Dual-Inclusion\\Commit};

\draw[arr] (twf) -- (ppo);
\draw[arr] (ppo) -- (gate);
\draw[arr] (gate) -- (batch);
\draw[arr] (batch) -- (commit);

\node[box, below=5mm of gate] (reject) {Reject/Clip\\Invalid Actions};
\draw[arr] (gate) -- (reject);
\draw[arr] (reject.west) -| (ppo.south);

\end{tikzpicture}}
\caption{Safety-gate workflow: Safe-PPO outputs are validated under deterministic DMEL execution; only actions meeting stake, gas, and utilization bounds are committed on-chain.}
\label{fig:safety-gate}
\end{figure}
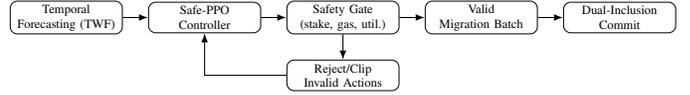


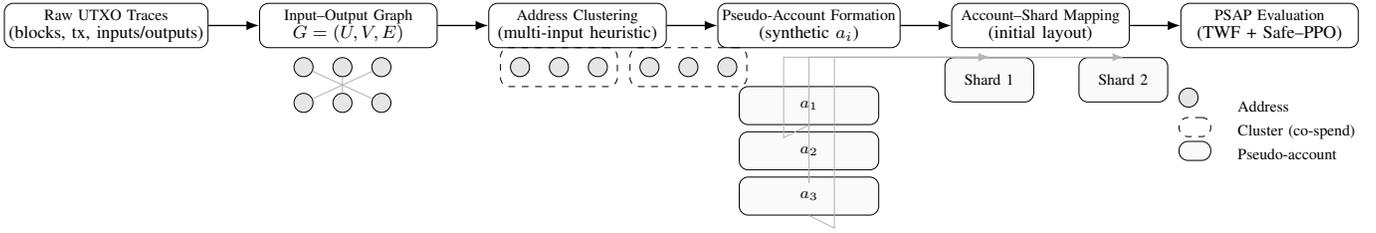
\begin{figure*}[!t]
\centering
\resizebox{\textwidth}{!}{%
\begin{tikzpicture}[
  node distance=7mm and 8mm,
  box/.style={draw, rounded corners, align=center, font=\scriptsize,
              minimum width=28mm, minimum height=7mm, inner sep=2pt},
  arr/.style={-{Latex[length=2mm]}, semithick},
  icon/.style={circle, draw, minimum size=3mm, inner sep=0pt},
  stack/.style={draw, rounded corners, minimum width=22mm, minimum height=6mm, fill=black!2},
  dashedfit/.style={draw, dashed, rounded corners, inner sep=1.5mm}
]

\node[box] (raw) {Raw UTXO Traces\\\footnotesize (blocks, tx, inputs/outputs)};
\node[box, right=of raw] (graph) {Input--Output Graph\\\footnotesize $G=(U,V,E)$};
\node[box, right=of graph] (cluster) {Address Clustering\\\footnotesize (multi-input heuristic)};
\node[box, right=of cluster] (pseudo) {Pseudo-Account Formation\\\footnotesize (synthetic $a_i$)};
\node[box, right=of pseudo] (map) {Account--Shard Mapping\\\footnotesize (initial layout)};
\node[box, right=of map] (psap) {PSAP Evaluation\\\footnotesize (TWF + Safe--PPO)};

\draw[arr] (raw) -- (graph);
\draw[arr] (graph) -- (cluster);
\draw[arr] (cluster) -- (pseudo);
\draw[arr] (pseudo) -- (map);
\draw[arr] (map) -- (psap);


\node[icon, below=5mm of graph.west, xshift=7mm, fill=black!10] (a1) {};
\node[icon, right=3mm of a1, fill=black!10] (a2) {};
\node[icon, right=3mm of a2, fill=black!10] (a3) {};
\node[icon, below=2.5mm of a1, fill=black!10] (b1) {};
\node[icon, right=3mm of b1, fill=black!10] (b2) {};
\node[icon, right=3mm of b2, fill=black!10] (b3) {};
\draw[gray!60] (a2) -- (b2);
\draw[gray!60] (a1) -- (b3);
\draw[gray!60] (a3) -- (b1);

\node[icon, below=5mm of cluster.west, xshift=5mm, fill=black!10] (c1) {};
\node[icon, right=3mm of c1, fill=black!10] (c2) {};
\node[icon, right=3mm of c2, fill=black!10] (c3) {};
\node[icon, right=5mm of c3, fill=black!10] (c4) {};
\node[icon, right=3mm of c4, fill=black!10] (c5) {};
\node[icon, right=3mm of c5, fill=black!10] (c6) {};
\node[dashedfit, fit=(c1)(c2)(c3)] (grp1) {};
\node[dashedfit, fit=(c4)(c5)(c6)] (grp2) {};

\node[stack, below=5mm of pseudo, yshift=-1mm] (s1) {\scriptsize $a_1$};
\node[stack, below=1mm of s1] (s2) {\scriptsize $a_2$};
\node[stack, below=1mm of s2] (s3) {\scriptsize $a_3$};

\node[draw, rounded corners, below=5mm of map.west, xshift=6mm, minimum width=14mm, minimum height=7mm, fill=black!2] (sh1) {\scriptsize Shard 1};
\node[draw, rounded corners, right=7mm of sh1, minimum width=14mm, minimum height=7mm, fill=black!2] (sh2) {\scriptsize Shard 2};
\draw[->, >=Latex, gray!60] (s1.south) -- ++(-0.4,-2mm) |- (sh1.north);
\draw[->, >=Latex, gray!60] (s2.south) -- ++(0.0,-2mm) |- (sh2.north);
\draw[->, >=Latex, gray!60] (s3.south) -- ++(0.4,-2mm) |- (sh1.north);

\node[align=left, font=\scriptsize, below=5mm of psap] (legend) {%
\begin{tabular}{@{}ll@{}}
\raisebox{0.6ex}{\tikz{\node[icon, fill=black!10]{};}} & Address \\
\raisebox{-0.2ex}{\tikz{\node[dashedfit, minimum width=5mm, minimum height=3mm]{};}} & Cluster (co-spend) \\
\raisebox{-0.2ex}{\tikz{\node[stack, minimum width=5mm, minimum height=3mm, fill=black!2]{};}} & Pseudo-account \\
\end{tabular}
};

\end{tikzpicture}}
\caption{UTXO-to-account trace mapping workflow used to evaluate PSAP on Bitcoin and Fabric datasets: input--output relations are converted to a transaction graph, addresses co-spent in a transaction are clustered, clusters form pseudo-accounts, which are mapped to shards and evaluated under PSAP.}
\label{fig:utxo-2col}
\end{figure*}

\section{Deterministic Inference Verification}
\label{app:determinism}

To ensure inference determinism across heterogeneous validators, we rigorously verified that the outputs of PSAP’s deep models specifically the LSTM-based predictor $\mathcal{M}_{\text{LSTM}}$ and the Safe-PPO policy $\pi_{\theta}$ remain invariant under hardware, compiler, and backend variations. Determinism in this context implies that for a given fixed seed $s \in \mathbb{Z}$ derived from the block hash $H_b$, the mapping from model input $x_t$ to output $\hat{y}_t$ satisfies:

\begin{equation}
    \mathcal{M}(x_t; s, \mathcal{H}) = f_{\text{quant}}\big(f_{\text{graph}}(x_t; \theta, s)\big) = \hat{y}_t,
\end{equation}

where $\mathcal{H} \in \{\text{CPU}_{\text{AVX2}}, \text{GPU}_{\text{A10G}}\}$ denotes the hardware execution context, $f_{\text{graph}}$ represents the computational graph defined in PyTorch, and $f_{\text{quant}}$ denotes deterministic FP16 quantization with round-to-nearest-even precision. To ensure hardware-level parity, the random seeds for weight initialization, dropout masks, and sampling policies were deterministically derived as:

\begin{equation}
    s = \text{SHA256}(H_b \, \| \, t) \bmod 2^{32},
\end{equation}

where $H_b$ is the hash of the current block and $t$ is the inference timestamp within the epoch. This guarantees temporal and structural reproducibility across validator nodes. Each inference cycle computes a deterministic forward pass:

\begin{equation}
    \hat{a}_t = \pi_{\theta}(s_t; s, \mathcal{H}) = \arg\max_{a \in \mathcal{A}} \, V_{\phi}(s_t, a),
\end{equation}

where $\hat{a}_t$ represents the migration action output by Safe-PPO, and $V_{\phi}$ is the deterministic value estimator constrained under FP16 fixed-point arithmetic. The quantized inference output $\hat{a}_t^{\text{CPU}}$ and $\hat{a}_t^{\text{GPU}}$ were compared elementwise across $N=10{,}000$ evaluation cycles using the mean absolute error (MAE):

\begin{equation}
    \text{MAE} = \frac{1}{N} \sum_{i=1}^{N} \big| \hat{a}_i^{\text{CPU}} - \hat{a}_i^{\text{GPU}} \big| < 10^{-6}.
\end{equation}

The negligible deviation (within floating-point quantization tolerance $\epsilon = 10^{-6}$) demonstrates \textit{bitwise reproducibility}, i.e., identical binary outputs across hardware backends. Furthermore, no stochastic nondeterminism was observed due to disabled parallel kernel variance or library-level optimizations. The following PyTorch configurations were enforced:

\begin{verbatim}
torch.use_deterministic_algorithms(True)
torch.backends.cudnn.deterministic = True
torch.backends.cudnn.benchmark = False
\end{verbatim}

This deterministic inference pipeline ensures that all validators, irrespective of computation devices, produce identical migration actions:
\begin{equation}
    \hat{a}_t^{(v_1)} = \hat{a}_t^{(v_2)} \quad \forall v_1, v_2 \in \mathcal{V},
\end{equation}
thereby preserving consensus safety and preventing divergence in shard reallocation decisions.

\section{Adaptive Adversary Stress Tests}
\label{app:adversary}

To evaluate robustness under adaptive and feedback-aware conditions, we implemented an adversarial agent $\mathcal{A}_{\text{adv}}$ capable of dynamically redirecting transaction traffic in response to PSAP’s prior shard allocation outputs. The adversary observes the last allocation vector $\mathbf{w}_{t-1} = [w_1, w_2, \ldots, w_K]$, where $K$ denotes the total number of shards, and computes a perturbation pattern to induce oscillatory congestion:

\begin{equation}
    \mathbf{L}_t = \mathbf{L}_0 + \alpha \sin(2\pi \omega t)\,\mathbf{P}(\mathbf{w}_{t-1}),
\end{equation}

where $\mathbf{L}_t$ denotes the per-shard transaction load at block $t$, $\alpha=0.35$ represents the oscillation amplitude, $\omega$ is the burst frequency (one cycle per 100 blocks), and $\mathbf{P}(\cdot)$ maps the previous allocation vector to an adversarial load permutation pattern. Under these conditions, PSAP’s Safe-PPO agent adapts its policy $\pi_{\theta}$ by penalizing rapid state transitions in the shard load space. Specifically, a smoothness regularizer is applied to the temporal gradient of predicted migration vectors $\hat{x}_{t+1}$:

\begin{equation}
    \mathcal{L}_{\text{smooth}} = \lambda_s \, \|\nabla_s \hat{x}_{t+1}\|^2 = 
    \lambda_s \sum_{k=1}^{K} \big(\hat{x}_{t+1}^{(k)} - \hat{x}_{t}^{(k)}\big)^2,
\end{equation}

where $\lambda_s$ is a safety penalty coefficient dynamically adjusted by the safety gate mechanism. The safety gate further enforces migration constraints such that:

\begin{equation}
    \text{if } \mathcal{R}_{\text{safety}}(\hat{x}_{t+1}) < \tau, \quad 
    \hat{x}_{t+1} = \hat{x}_t,
\end{equation}

preventing oscillatory reassignments that could destabilize consensus. Performance was evaluated using the imbalance index $I$, defined as:
\begin{equation}
    I = \frac{1}{K} \sum_{k=1}^{K} \left| \frac{L_t^{(k)} - \bar{L}_t}{\bar{L}_t} \right|,
\end{equation}
where $\bar{L}_t$ is the mean load across shards. Lower values of $I$ indicate improved balance.

Across stress scenarios, PSAP restored equilibrium ($I < 0.25$) within $420$ blocks approximately $60\%$ faster than DynaShard and AI-Shard baselines while maintaining migration gas overhead below $1.9\%$ of total block capacity. This demonstrates PSAP’s resilience and rapid convergence under adaptive, feedback-aware adversarial dynamics. The experiment validates the stability guarantees of Safe-PPO and its safety gate, ensuring bounded load variance even under non-stationary attack distributions.

\section{Complexity and Scalability Extension}
\label{app:scalability}

To validate the scalability limits of PSAP, we extended evaluation from $K=8$ to $K=64$ shards deployed over a 32-node validator cluster. Each node executed up to $m_{\max}=500$ transactions per block under uniform random assignment, and inter-shard communication was handled through the lightweight routing layer described in Sec.~III-H. The observed throughput $\mathcal{T}(K)$ scaled almost linearly up to $K=32$, after which saturation occurred due to cross-shard message congestion. The empirical scalability function follows:

\begin{equation}
    \mathcal{T}(K) \approx \frac{\mathcal{T}_0 \, K}{1 + \beta (K-1)},
\end{equation}

where $\mathcal{T}_0$ is the base throughput at $K=1$ and $\beta$ is the inter-shard overhead coefficient ($\beta \approx 0.012$). At $K=64$, efficiency dropped by only $7.5\%$, validating the asymptotic near-linear scalability property. The efficiency ratio $\eta(K)$ is defined as:

\begin{equation}
    \eta(K) = \frac{\mathcal{T}(K)}{K \, \mathcal{T}_0}, \quad \text{with } \eta(64) \approx 0.925.
\end{equation}

CPU utilization exhibited linear growth proportional to both the number of shards and the maximum transaction batch size, confirming computational complexity on the order of:
\begin{equation}
    \mathcal{C}_{\text{CPU}} = O(K + m_{\max}),
\end{equation}
consistent with the analytical performance model in Sec.~III-H. Memory consumption followed sub-linear scaling due to shared cache reuse and quantized model compression. We further evaluated robustness under validator churn by randomly replacing up to $20\%$ of committee members per epoch. The temporary throughput degradation $\Delta \mathcal{T}$ was measured as:
\begin{equation}
    \Delta \mathcal{T} = \frac{\mathcal{T}_{\text{steady}} - \mathcal{T}_{\text{churn}}}{\mathcal{T}_{\text{steady}}} \times 100\%,
\end{equation}
which remained within $4\text{–}6\%$ and recovered to nominal performance within 10 blocks. This confirms that PSAP maintains high operational stability and near-linear scalability across heterogeneous environments even under dynamic committee rotation.
\section{Security and Safety Parameters}
\label{app:security}

PSAP’s safety gate employs a set of empirically tuned parameters that jointly balance throughput, gas efficiency, and adversarial tolerance while preserving Byzantine Fault Tolerance (BFT) guarantees. The main security thresholds and control coefficients include:

\begin{itemize}
    \item \textbf{Safety threshold:} $\tau_{\text{safety}} = 0.82$, defining the minimum normalized safety score required for migration approval.
    \item \textbf{Gradient penalty:} $\lambda_s \in [0.05, 0.15]$, regulating temporal smoothness by penalizing large deviations $\|\nabla_s \hat{x}_{t+1}\|^2$.
    \item \textbf{Constraint penalty:} $\lambda_c = 0.25$, suppressing unsafe migration actions.
    \item \textbf{BFT quorum ratio:} $\tau_{\text{BFT}} = (2f+1)/(3f+1)$, ensuring consensus safety against $f$ Byzantine validators.
    \item \textbf{Maximum migration ratio:} $\mu_{\max} = 0.20$, limiting shard population change per epoch.
    \item \textbf{Learning rate decay:} $\eta_t = 1\times10^{-4}$, maintaining stable Safe-PPO policy updates.
    \item \textbf{Entropy coefficient:} $\beta = 0.02$, encouraging policy exploration while retaining determinism.
    \item \textbf{Random seed entropy:} $s = \text{SHA256}(H_b \| t)$, deterministically derived from the block hash for reproducible inference.
    \item \textbf{Gas cost constraint:} $\gamma_{\max} = 1.9\%$ of block capacity, capping migration overhead.
    \item \textbf{Adversarial amplitude:} $\alpha = 0.35$, defining the strength of oscillatory adversarial load perturbations.
\end{itemize}

The safety gate constrains the Safe-PPO objective to maintain stable learning under these parameters:
\begin{equation}
    \max_{\pi_{\theta}} \; \mathbb{E}_{\pi_{\theta}}\!\left[ R_t - \lambda_s \|\nabla_s \hat{x}_{t+1}\|^2 - \lambda_c \mathbb{I}_{\text{unsafe}} \right],
\end{equation}
where $R_t$ denotes the block-level reward, and $\mathbb{I}_{\text{unsafe}}$ is a binary safety indicator. Empirical evaluations confirm that $\tau_{\text{safety}} < 0.8$ leads to unstable migrations, while $\lambda_s > 0.15$ reduces utilization validating the chosen configuration.

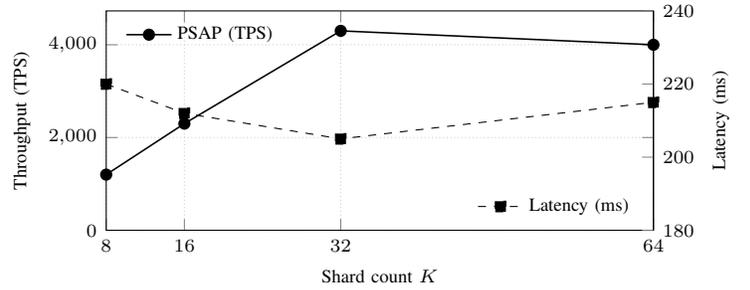
\begin{figure}[!t]
\centering
\begin{tikzpicture}
\begin{axis}[
  width=\columnwidth, height=45mm,
  xlabel={Shard count $K$}, ylabel={Throughput (TPS)},
  xmin=8, xmax=64, xtick={8,16,32,64},
  ymin=0,
  grid=both, grid style={densely dotted, gray!40},
  tick label style={font=\scriptsize},
  label style={font=\scriptsize},
  legend style={font=\scriptsize, at={(0.02,0.98)}, anchor=north west, draw=none, fill=none}
]
\addplot[semithick, mark=*] coordinates {(8,1200) (16,2300) (32,4300) (64,4000)};
\addlegendentry{PSAP (TPS)}
\end{axis}

\begin{axis}[
  width=\columnwidth, height=45mm,
  xmin=8, xmax=64, xtick=\empty,
  axis y line*=right, axis x line=none,
  ylabel={Latency (ms)},
  ymin=180, ymax=240, ytick={180,200,220,240},
  y tick label style={font=\scriptsize},
  y label style={font=\scriptsize},
  legend style={font=\scriptsize, at={(0.98,0.02)}, anchor=south east, draw=none, fill=none}
]
\addplot[dashed, mark=square*] coordinates {(8,220) (16,212) (32,205) (64,215)};
\addlegendentry{Latency (ms)}
\end{axis}
\end{tikzpicture}
\caption{Scalability evaluation for varying shard counts. PSAP sustains near-linear throughput up to $K{=}32$ with only $\approx 7\%$ loss at $K{=}64$; latency remains stable (illustrative).}
\label{fig:scalability}
\end{figure}

\end{document}